\documentclass[reprint,aps,prb,amsmath,amssymb,superscriptaddress]{revtex4-1}
\usepackage{graphicx}
\usepackage{hyperref}
\hypersetup{colorlinks=true, linkcolor=blue, citecolor=blue, urlcolor=blue}

\renewcommand{\v}[1]{\mathbf{#1}}        
\newcommand{\gv}[1]{\mbox{\boldmath$ #1 $}}
\newcommand{\uv}[1]{\mathbf{\hat{#1}}}

\renewcommand{\d}[0]{\ensuremath{\operatorname{d}\!}}
 
\newcommand{\td}[2]{\frac{\d #1}{\d #2}}        
        
\newcommand{\pd}[2]{\frac{\partial #1}{\partial #2}}

\let\f=\frac

\renewcommand{\l}[0]{\left}
\renewcommand{\r}[0]{\right}
\newcommand{\fp}[2]{\l( \f{#1}{#2} \r)}

\newcommand{\avg}[1]{\langle #1 \rangle}
\newcommand{\Avg}[1]{\l< #1 \r>}

\newcommand{\dmat}[4]{\begin{vmatrix} #1 & #2 \\ #3 & #4 \end{vmatrix}}
\newcommand{\Dmat}[9]{\begin{vmatrix} #1 & #2 & #3 \\ #4 & #5 & #6 \\ #7 & #8 & #9\end{vmatrix}}

\newcommand{\ie}[0]{\textit{i.e.}}
\newcommand{\eg}[0]{\textit{e.g.}}

\renewcommand{\t}[1]{\text{#1}}

\newcommand{\Ang}[0]{\, \mathring{\mathrm{A}}}

\newcommand{\kB}[0]{k_\t{B}}
\newcommand{\K}[0]{\, \t{K}}
\newcommand{\ns}[0]{\, \t{ns}}

\begin{document}
  \title{Step free energies at faceted solid surfaces: Theory and atomistic calculations for steps on the Cu(111) surface}
  \author{Rodrigo Freitas}
  \email{rodrigof@berkeley.edu}
  \affiliation{Department of Materials Science and Engineering, University of California, Berkeley, CA 94720, USA} 
  \affiliation{Lawrence Livermore National Laboratory, Livermore, CA 94550, USA}
  \author{Timofey Frolov}
  \affiliation{Lawrence Livermore National Laboratory, Livermore, CA 94550, USA}
  \author{Mark Asta}
  \affiliation{Department of Materials Science and Engineering, University of California, Berkeley, CA 94720, USA} 
  \date{\today}
  \begin{abstract}
    A theory for the thermodynamic properties of steps on faceted crystalline surfaces is presented. The formalism leads to the definition of step excess quantities, including an excess step stress that is the step analogy of surface stress. The approach is used to develop a relationship between the temperature dependence of the step free energy ($\gamma^\t{st}$) and step excess quantities for energy and stress that can be readily calculated by atomistic simulations. We demonstrate the application of this formalism in thermodynamic-integration (TI) calculations of the step free energy, based on molecular-dynamics simulations, considering $\avg{110}$ steps on the $\{111\}$ surface of a classical potential model for elemental Cu. In this application we employ the Frenkel-Ladd approach to compute the reference value of $\gamma^\t{st}$ for the TI calculations. Calculated results for excess energy and stress show relatively weak temperature dependencies up to a homologous temperature of approximately 0.6, above which these quantities increase strongly and the step stress becomes more isotropic. From the calculated excess quantities we compute $\gamma^\t{st}$ over the temperature range from zero up to the melting point ($T_\t{m}$). We find that $\gamma^\t{st}$ remains finite up to $T_\t{m}$, indicating the absence of a roughening temperature for this $\{111\}$ surface facet, but decreases by roughly fifty percent from the zero-temperature value. The strongest temperature dependence occurs above homologous temperatures of approximately 0.6, where the step becomes configurationally disordered due to the formation of point defects and appreciable capillary fluctuations. 
  \end{abstract}

  \maketitle

  \section{\label{sec:introduction} Introduction}
    In theories of crystal morphologies and growth kinetics, a property of fundamental importance is the step free energy, $\gamma^\t{st}$, \ie, the excess free energy of a step line defect on a faceted solid-liquid interface or crystal surface \cite{williams_review,williams}. The magnitude of $\gamma^\t{st}$ controls the depth of the cusp in the interfacial free energy versus orientation plot for faceted interfaces, and this property is thus fundamental in determining the equilibrium crystal shape \cite{einstein_review}. The step free energy also plays an important role in governing crystallization kinetics from the melt or vapor \cite{chris_orme}, by controlling the magnitude of the barrier to island nucleation in the growth of a faceted surface or interface. The step free energy can depend strongly on temperature, and this dependence ultimately leads to the vanishing of $\gamma^\t{st}$ above the thermodynamic roughening temperature \cite{faceting, *[{Erratum: }]faceting_erratum}. Below the roughening transition, steps with free energies that are low relative to the thermal energy will display pronounced capillary fluctuations, which have important consequences for their kinetic properties, bunching instabilities \cite{bunching}, and morphologies \cite{bartelt,wandering}.
    
    Despite the importance of $\gamma^\t{st}$ described above, measurements of this quantity remain relatively rare. Further, reported values are often available only for a fixed value of the temperature \cite{exp_1,exp_3,exp_4,exp_5} and measurements over a wide temperature range have been undertaken in few systems \cite{exp_2,exp_6}. As a consequence, knowledge of the nature of the temperature dependence of step free energies and understanding of the microscopic factors that underlie it remain incomplete. This situation presents a challenge for the development and application of quantitative mesoscale theories in studies of faceted crystal growth phenomena in real systems, and robust methods for the direct calculation of temperature-dependent step free energies from atomic-scale simulations are thus of fundamental interest. In the present paper we present a thermodynamic formalism that relates the temperature dependence of $\gamma^\t{st}$ on faceted crystal surfaces to excess quantities that can be computed directly by atomistic simulations. This formalism provides a framework for the calculation of $\gamma^\t{st}$ as a function of temperature through the thermodynamic integration of an appropriate adsorption equation.
    
    The remainder of this paper is organized as follows. In Sec.~\ref{sec:steps_theory} the thermodynamic formalism is introduced, and the relevant step excess quantities and other fundamental thermodynamic equations are defined. In Sec.~\ref{sec:TI_theory} we demonstrate how the thermodynamic formalism can be combined with the calculation of a reference step free energy at low temperatures by the Frenkel-Ladd method \cite{frenkel_ladd,freitas_fl}, to compute $\gamma^\t{st}$ up to high temperatures, accounting for contributions arising from the formation of surface point defects and capillary fluctuations. In Secs.~\ref{sec:simulations} and \ref{sec:results} we present simulation details and results, respectively, of an application of the equations derived to the calculation of the free energy of $\avg{110}$ steps present on the $\{111\}$ surface of face-centered-cubic copper using molecular dynamics simulations. In Sec.~\ref{sec:results} we also compare the step free energy obtained here to experimentally measured \cite{exp_4} and first-principles-calculated \cite{dft_1,dft_2,dft_3} results available in the literature.  Finally, in Sec.~\ref{sec:summary} we summarize the main findings, and discuss applications of the formalism presented in this work more generally.

  \section{\label{sec:steps_theory} Thermodynamic theory of surface steps}
    \subsection{\label{sec:step_excess} Step excess quantities}
\begin{figure}
        \centering
        \includegraphics[width=0.48\textwidth]{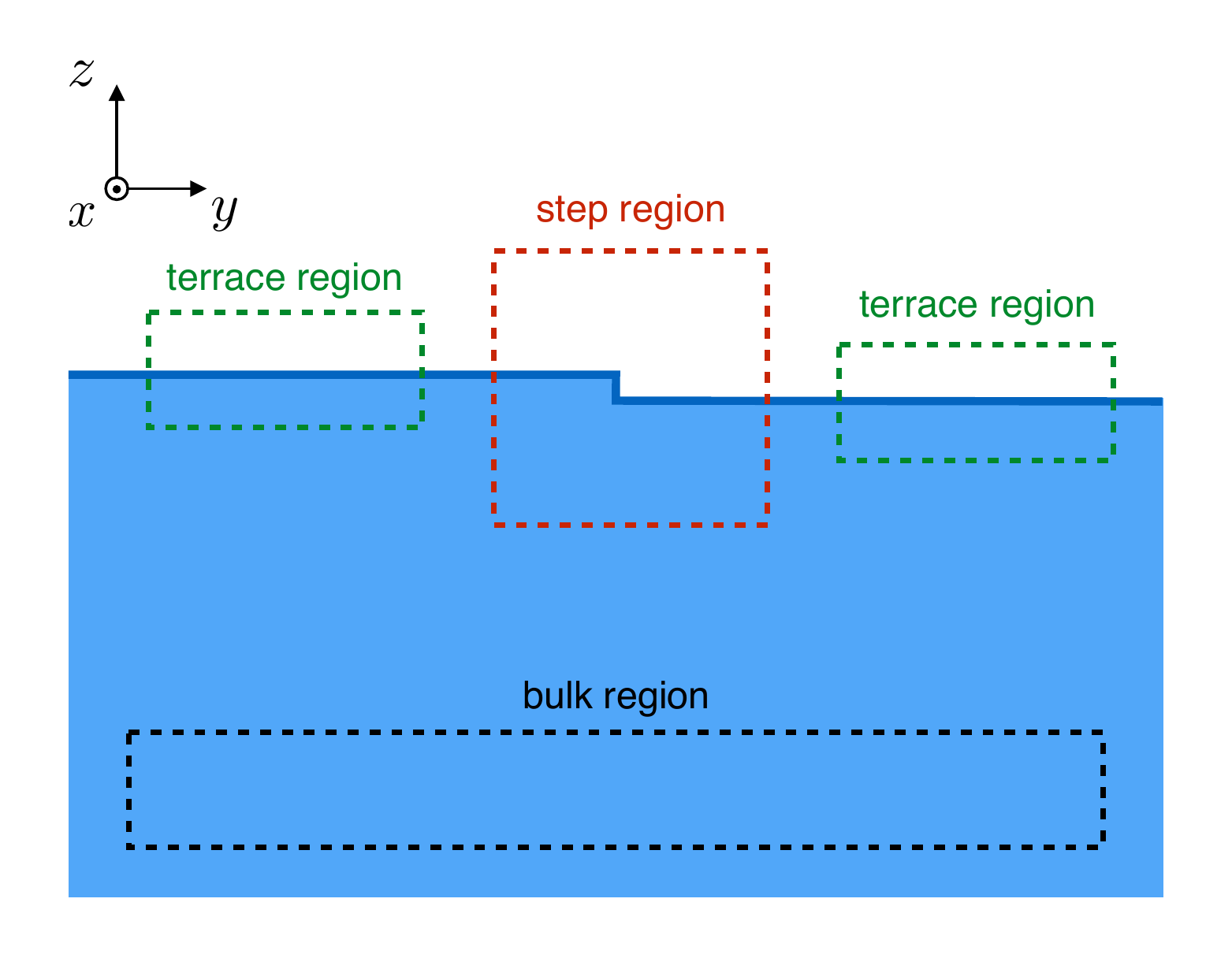}
        \caption{\label{fig:step_config} Schematic illustration of a thermodynamic system consisting of a homogeneous solid and a stepped surface. The solid is infinite along the $\pm \uv{x}$, $\pm \uv{y}$, and $-\uv{z}$ directions. The surface normal is $\uv{z}$ and the average step direction is along $\uv{x}$.}
      \end{figure}

      Consider a thermodynamic system that consists of a homogeneous solid with a stepped surface, where the step separates two flat surface terraces as shown in Fig.~\ref{fig:step_config}. Both terraces have the same structure and thermodynamic properties, while the surface around the step has properties that are different from those of the flat terraces. The step region, terraces, and bulk are in thermodynamic equilibrium with each other. We assume that atoms can migrate by diffusion between the bulk and the surface regions, allowing the concentration of point defects to vary everywhere in the system, in a way required to maintain equilibrium. We also assume that atoms can attach and detach from the step, so the system is in equilibrium with an infinite source and sink of atoms.
    
      The properties of the bulk far away from the surface region and the properties of the flat terraces far away from the step region are well described by standard bulk and interfacial thermodynamic relations \cite{gibbs,cahn}. In this section we will address the thermodynamic properties of the steps on the crystalline surface. Consider an imaginary region that contains a finite segment of the step, as shown in Fig.~\ref{fig:step_config}. The lower boundary of this region is located inside the homogeneous part of the bulk crystal, while the side boundaries parallel and perpendicular to the step line cross the system surface normal to the terraces. The latter condition is important because it defines the total surface area inside the region.
    
      The extensive thermodynamic properties of the step region depend on its dimensions, since the enclosed system is not homogeneous. We postulate that the total energy of the region is a function of the following extensive and intensive variables:
      \begin{equation}
        \label{eq:E_st_def}
        E^\t{st} \equiv E^\t{st} (S^\t{st}, N^\t{st}, A^\t{st}, L, \varepsilon_{ij}).
      \end{equation}
      The superscript ``st'' refers to variables of the step region shown in Fig.~\ref{fig:step_config}: $S^\t{st}$ is the entropy, $N^\t{st}$ is the number of atoms, $A^\t{st}$ is the surface area enclosed by the region, $L$ is the step length, and $\varepsilon_{ij}$ are the lateral components of strain, with $i=x,y$ and $j=x,y$. The lateral components of strain in Eq.~\eqref{eq:E_st_def} correspond to the macroscopic strain in the homogeneous bulk lattice far away from the step, not to be confused with the local inhomogeneous strain around the step. As illustrated in Fig.~\ref{fig:step_config}, the coordinate system is chosen such that $\uv{z}$ is normal to the terraces, while $\uv{x}$ and $\uv{y}$ are parallel and normal to the step line, respectively.
    
      Consider a variation when the physical state of the system is fixed and we extend the boundaries of the region from zero to some finite values. Assuming that $E^\t{st}$ is a homogeneous function of degree one with respect to $S^\t{st}$, $N^\t{st}$, $A^\t{st}$, and $L$ we obtain
      \begin{equation}
        \label{eq:E_st}
        E^\t{st} = TS^\t{st} + \mu N^\t{st} + \gamma A^\t{st} + \gamma^\t{st} L,
      \end{equation}
      where $T$ is the temperature, $\mu$ is the chemical potential, $\gamma$ is the surface free energy per unit area, and $\gamma^\t{st}$ is the step free energy per unit step length. Note that by definition $\gamma$ is the property of the terrace uninfluenced by the surface step. We assume that the external pressure is zero since the solid is in contact with vacuum.
    
      At this point a comment should be made about the meaning of the quantities introduced above. The current thermodynamic treatment is focused on steps on solid surfaces, and it is well known that such steps produce long-range elastic fields \cite{marchenko,srolovitz,muller}. As a result, both terraces and the bulk crystal are strictly speaking inhomogeneous in the entire system. Equation \eqref{eq:E_st} can still be used to describe the system if $\gamma$ and $\gamma^\t{st}$ are understood as the properties of the terraces and the step in the limit when the system size goes to infinity. In other words, even though the inhomogeneity due to the strain fields induced by the step can extend far away from the step line, its total contribution to the energy of the system is finite. This will be demonstrated using atomistic simulations in Sec.~\ref{sec:simulations} of this study. This property of surface steps should be contrasted with the case of lattice dislocations, which are also line defects. Different from steps, the elastic contribution to the total energy of a dislocation diverges with the system size \cite{hirth} and Eq.~\eqref{eq:E_st} would not apply.
    
      The amount of bulk and terrace inside the step region at this point is arbitrary, and hence quantities in Eq.~\eqref{eq:E_st} depend on the choice of the step region. In order to define the step excess quantities we need to subtract the bulk and terrace contributions from the quantities of the step region in Fig.~\ref{fig:step_config}. To this end we write equations analogous to Eq.~\eqref{eq:E_st} for the terrace and bulk regions shown in Fig.~\ref{fig:step_config}:
      \begin{align}
        \label{eq:E_t}
        E^\t{t} & = TS^\t{t} + \mu N^\t{t} + \gamma A^\t{t} \\
        \label{eq:E_b}
        E^\t{b} & = TS^\t{b} + \mu N^\t{b},
      \end{align}
      where superscripts ``t'' and ``b'' refer to terrace and bulk respectively. These two regions are located sufficiently far away from the step that their extensive properties are not affected by it. The terrace region includes the surface as well as a portion of the homogeneous bulk phase, while the bulk region is unaffected by the surface. Solving the system of equations given by Eqs.~\eqref{eq:E_st}, \eqref{eq:E_t}, and \eqref{eq:E_b} using Cramer's rule, we obtain an expression for step free energy $\gamma^\t{st}$:
      \begin{equation}
        \label{eq:gamma_excess}
        \gamma^\t{st} L = [E - TS - \mu N - \gamma A]_{XY}
      \end{equation}
      where $X$ and $Y$ are any of the extensive quantities $S$, $N$ or $A$. Terms $[Z]_{XY}$ are Cahn's determinants and are calculated as the ratio of two determinants \cite{cahn}
      \begin{equation}
        \label{eq:excess_definition}
        [Z]_{XY} = \f{\Dmat{Z^\t{st}}{X^\t{st}}{Y^\t{st}}{Z^\t{t}}{X^\t{t}}{Y^\t{t}}{Z^\t{b}}{X^\t{b}}{Y^\t{b}}}{\dmat{X^\t{t}}{Y^\t{t}}{X^\t{b}}{Y^\t{b}}}.
      \end{equation}
      The first row of the numerator contains extensive thermodynamic quantities of the region containing the step, while the second and the third rows contain properties of regions enclosing the terrace and the bulk, respectively. According to properties of determinants if any two columns are equal, the determinant is zero:
      \begin{equation}
        \label{eq:det_prop}
        [X]_{XY} = [Y]_{XY} = 0.
      \end{equation}
      Thus, two terms in Eq.~\eqref{eq:gamma_excess} automatically vanish.
    
      The quantity $[Z]_{XY}$ has the meaning of the excess property of a step when the region with the step has the same amount of $X$ and $Y$ as the terrace and the bulk regions combined. The excess quantities generally depend on the choice of the extensive variables $X$ and $Y$. On the other hand, Eq.~\eqref{eq:gamma_excess} shows that all different choices of $X$ and $Y$ result in the same excess amount of $\gamma^\t{st} L$.
    
      Considering a particular example when $X = A$ and $Y = N$, we obtain
      \begin{equation}
        \label{eq:gamma_st}
        \gamma^\t{st} L = [E]_{AN} - T[S]_{AN}.
      \end{equation}
      The step excess quantities $[E]_{AN}$ and $[S]_{AN}$ are the excess energy and entropy when the step region has the same surface area and number of atoms as terrace and bulk regions combined. The choice of $A$ as one of the extensive variables means that the excess area of a step is zero. The step is represented as a dividing line on the surface and the properties of terraces are extended all the way to this line. Using this representation, step excess quantities can be formulated in a manner similar to the Gibbs dividing surface construction for interfaces. On the other hand, the derivation that uses Cahn's determinants provides expressions for excess quantities that are more general. The ability to choose different definitions can be useful in applications because some excess quantities are more accessible than others to measurements or calculations. In Sec.~\ref{sec:simulations} we describe how several step excess quantities can be calculated directly from atomistic simulations by making use of the flexibility provided by Cahn's determinants.
    
    \subsection{\label{sec:step_adsorp} Adsorption equation} 
      In the previous section we derived an expression for the step free energy and other excess quantities. We are now in a position to derive an equation that describes how $\gamma^\t{st}$ changes with temperature and mechanical deformation, namely the adsorption equation. Consider a variation of state when the system exchanges heat and does mechanical work. For the region containing the step the change in total energy is given by
        \begin{equation}
          \d{E^\t{st}} = T\d{S^\t{st}} + \mu \d{N^\t{st}} + \sum_{i,j}^{x, y} \sigma_{ij}^\t{st} V^\t{st} \d{\varepsilon_{ij}}, 
          \label{eq:dE_st}
        \end{equation}
      where $\sigma^\t{st}_{ij}$ is the stress tensor and $V^\t{st}$ is the volume. The product $\sigma^\t{st}_{ij} V^\t{st}$ is defined as the derivative of $E^\t{st}$ with respect to elastic deformation $\varepsilon_{ij}$. Equation \eqref{eq:dE_st} assumes that the surface area changes due to elastic deformation of the lattice and not by incorporation of new lattice units. At the same time the number of atoms in the region and the relative areas of the terraces can change by diffusion and attachment of atoms to the step. The conditions for mechanical equilibrium between the system and the vacuum \cite{landau,muller} require that $\sigma^\t{st}_{iz} = 0$ for $i = x,\, y,\, \t{or}\, z$. Thus, all summations involving the stress tensor are over the $x$ and $y$ indices only.
    
      Performing a Legendre transformation on terms containing entropy and number of particles we obtain from Eq.~\eqref{eq:dE_st}
      \begin{align}
        \label{eq:legendre_trans}
        \d{(E^\t{st}} - T S^\t{st} - \mu N^\t{st}) = -S^\t{st} \d{T} & - N^\t{st} \d{\mu} \\
        & + \sum_{i,j}^{x,y} \sigma_{ij}^\t{st} V^\t{st} \d{\varepsilon_{ij}}. \nonumber
      \end{align}
      Combining Eqs.~\eqref{eq:E_st} and (\ref{eq:legendre_trans}) we obtain
      \begin{align}
        \label{eq:dgamma_st_1}
        \d{(\gamma^\t{st}L)} = - S^\t{st} \d{T} & - N^\t{st} \d{\mu} \\
          & - A^\t{st} \d{\gamma}+ \sum_{i,j}^{x,y} ( \sigma_{ij}^\t{st} V^\t{st} - \delta_{ij} \gamma A^\t{st}) \d{\varepsilon_{ij}} . \nonumber
      \end{align}
      The intensive variables on the right-hand side in Eq.~\eqref{eq:dgamma_st_1} are not independent since equations similar to Eq.~\eqref{eq:dgamma_st_1} for the terrace and bulk regions impose additional constrains. For the terrace we have \cite{tim}
      \begin{align}
        \label{eq:dgamma_st_2}
        0 = - S^\t{t} \d{T} & - N^\t{t} \d{\mu} \\
        & - A^\t{t} \d{\gamma} + \sum_{i,j}^{x,y} ( \sigma_{ij}^\t{t} V^\t{t} - \delta_{ij} \gamma A^\t{t} ) \d{\varepsilon_{ij}}, \nonumber
      \end{align}
      while the Gibbs-Duhem equation for the bulk reads
      \begin{equation}
        \label{eq:dgamma_st_3}
        0 = -S^\t{b} \d{T} - N^\t{b} \d{\mu} + \sum_{i,j}^{x,y} \sigma_{ij}^\t{b} V^\t{b} \d{\varepsilon_{ij}}.
      \end{equation}
      Solving Eqs.~\eqref{eq:dgamma_st_1}, \eqref{eq:dgamma_st_2}, and \eqref{eq:dgamma_st_3} using Cramer's rule \cite{cahn}, we obtain the adsorption equation for steps
      \begin{align}
        \label{eq:adsorption}
        \d{(\gamma^\t{st}L)} = - [S]_{XY}\d{T} & - [N]_{XY} \d{\mu} \\
        - [A]_{XY} \d{\gamma} & + \sum_{i,j}^{x,y} \l[\sigma_{ij} V - \delta_{ij} \gamma A\r]_{XY} \d{\varepsilon_{ij}}, \nonumber
      \end{align}
      where $X$ and $Y$ are any of the extensive quantities $S$, $N$, $A$, or $(\sigma_{ij} V - \delta_{ij} \gamma A)$. Notice that the coefficients of the differentials in Eq.~\eqref{eq:adsorption} are the step excess quantities introduced earlier in Eq.~\eqref{eq:excess_definition}, and are independent of the particular choice of the regions illustrated in Fig.~\ref{fig:step_config}. Due to the property of determinants in Eq.~\eqref{eq:det_prop}, two terms in the adsorption equation can be eliminated by specifying $X$ and $Y$, leaving only independent variables. The number of variables should coincide with the number of degrees of freedom available to the system. Consider the same example given in Sec.~\ref{sec:step_excess}, where we choose $X$ and $Y$ equal to $A$ and $N$. In this case the four possible variations are changes in temperature and deformation described by strains $\varepsilon_{xx}$, $\varepsilon_{yy}$, and $\varepsilon_{xy}$. It is natural to have the step free energy be a function of these variables. The differential of surface free energy $\gamma$ that appears in Eq.~\eqref{eq:adsorption} is an unusual variable to describe the changes in the thermodynamic state of the step. While such an exotic form of the adsorption equation can be formulated and is consistent with the Gibbs phase rule, in most practical cases it is more convenient to eliminate this term by specifying $X = A$.
    
    \subsection{Step stress}
      Equation \eqref{eq:adsorption} introduces a new excess property in addition to the quantities that appeared in Eq.~\eqref{eq:gamma_excess}. The last term in Eq.~\eqref{eq:adsorption} describes changes in $\gamma^\t{st} L$ due to elastic deformations and defines the step excess stress as 
      \begin{equation}
        \label{eq:Shuttleworth_step}
        [\tau_{ij}]_{XY} \equiv \f{1}{L} \pd{\l( \gamma^\t{st}L \r)}{\varepsilon_{ij}} = \f{1}{L} \l[\sigma_{ij} V - \delta_{ij} \gamma A \r]_{XY},
      \end{equation}
      where $i = x, y$ and $j = x, y$. $[\tau_{ij}]_{XY}$ is a quantity with units of energy per length that represents the additional force exerted on the perimeter of the stepped surface due to the presence of the step. Different from $\gamma^\t{st}$, the step excess stress $[\tau_{ij}]_{XY}$ is not a unique quantity: it is a direct consequence of the derived adsorption equation that one can introduce several valid step excess stresses by specifying different extensive properties $X$ and $Y$. Notice that by the derivation above $[\gv{\tau}]$ is a second rank tensor, not a scalar like step free energy; hence it has nonzero components parallel and normal to the step line \cite{li}. 
    
      Differentiating the product $\gamma^\t{st}L$ in Eq.~\eqref{eq:adsorption} and using $\d{L} = \sum_{i,j}^{x,y} \delta_{ix} \delta_{jx} L \d{\varepsilon_{ij}}$ we obtain the intensive form of the adsorption equation:
      \begin{align}
        \label{eq:adsorption_intensive}
        \d{\gamma^\t{st}} = & - \f{[S]_{XY}}{L} \d{T} - \f{[N]_{XY}}{L} \d{\mu} \\
        &  - \f{[A]_{XY}}{L} \d{\gamma} + \sum_{i,j}^{x,y} ([\tau_{ij}]_{XY} - \delta_{ix} \delta_{jx} \gamma^\t{st} ) \d{\varepsilon_{ij}}\nonumber,
      \end{align}
      where the differential coefficients are the step excess quantities per unit step length. From Eq.~\eqref{eq:adsorption_intensive} we can now obtain the relation between $[\tau_{ij}]_{XY}$ and $\gamma^\t{st}$:
      \begin{equation}
        \label{eq:Shuttleworth_step_2}
        [\tau_{ij}]_{XY} = \delta_{ix} \delta_{jx} \gamma^\t{st} + \pd{\gamma^\t{st}}{\varepsilon_{ij}}.
      \end{equation}
      Equations \eqref{eq:Shuttleworth_step} and \eqref{eq:Shuttleworth_step_2} are the step analogs of the stress equations for solid surfaces \cite{shuttleworth,tim,Frolov2012a}.  They are a direct consequence of the derived adsorption equation, Eq.~\eqref{eq:adsorption}, and give a recipe for how $[\tau_{ij}]_{XY}$ can be calculated as an excess property using the determinant formalism. 
    
      Consider the example discussed earlier (Secs.~\ref{sec:step_excess} and \ref{sec:step_adsorp}) when $X = A$ and $Y = N$. This choice of extensive variables eliminates differentials of chemical potential $\mu$ and surface free energy $\gamma$, leaving only independent variations with temperature and deformation:
      \begin{align}
        \label{eq:dgamma_AN}
        \d{\gamma^\t{st}} = - \f{[S]_{AN}}{L} \d{T} + \sum_{i,j}^{x,y} ([\tau_{ij}]_{AN} - \delta_{ix} \delta_{jx} \gamma^\t{st} ) \d{\varepsilon_{ij}}.
      \end{align}
      The second term in Eq.~\eqref{eq:dgamma_AN} describes how the step free energy changes when the surface is deformed at constant temperature. Notice that during such a process the chemical potential $\mu$ and free energy of the terraces $\gamma$ are not constant. Equation \eqref{eq:dgamma_AN} defines a particular step excess stress given by
      \begin{equation}
        [\tau_{ij}]_{AN} = \f{1}{L} \l[ \sigma_{ij} V \r]_{AN}.
        \label{eq:step_tau_NA}
      \end{equation}
      The components of this stress tensor have been calculated in the present work from atomistic simulations, and the magnitudes of this quantity will be presented in Sec.~\ref{sec:results} below.

  \section{\label{sec:TI_theory} Thermodynamic integration formalism}
    In this section we describe how the equations derived in Sec.~\ref{sec:steps_theory} provide a framework for a thermodynamic-integration approach to computing the temperature dependence of the step free energy $\gamma^\t{st}$ by atomistic simulations. We also demonstrate how the absolute free energy of the step can be derived at low temperatures (\ie, where the concentration of kinks and surface adatoms are sufficiently low that we can neglect their contribution to the free energy) using the Frenkel-Ladd \cite{frenkel_ladd} method, to provide a reference value in the thermodynamic integration approach. The combination of these two methods provides a general framework for the calculation of step free energies over a wide temperature range, accounting naturally for vibrational and configurational disorder.

    \subsection{\label{sec:TI_for_steps} Gibbs-Helmholtz relation for step free-energy integration}
      The temperature dependence of the step free energy can be obtained by directly integrating $\d{(\gamma^\t{st}L)}$, given in Eq.~\eqref{eq:adsorption}, along a reversible thermodynamic trajectory. However, in many applications the calculation of the excess entropy $[S]_{XY}$ can be challenging. Fortunately, we can avoid the explicit calculation of $[S]_{XY}$ by integrating $\d{(\gamma^\t{st}L/T)}$ instead of Eq.~\eqref{eq:adsorption}. We can compute $\d{(\gamma^\t{st}L/T)}$ explicitly by combining Eq.~\eqref{eq:gamma_excess} and \eqref{eq:adsorption}:
      \begin{align}
        \d{\fp{\gamma^\t{st}L}{T}} = & - \f{[E - \mu N -\gamma A]_{XY}}{T^2} \d{T} - \f{[N]_{XY}}{T} \d{\mu} \nonumber \\
         & - \f{[A]_{XY}}{T} \d{\gamma} + \sum_{i,j}^{x,y} \f{[\tau_{ij}]_{XY} L}{T} \d{\varepsilon_{ij}},
        \label{eq:GH_steps}
      \end{align}
      where $[\tau_{ij}]_{XY}$ depends on the choice of the $X$ and $Y$ variables. Equation \eqref{eq:GH_steps} is the surface step analog of the Gibbs-Helmholtz equation from bulk thermodynamics. A similar equation for interfaces was derived previously \cite{Frolov09jcp}, and was demonstrated to be efficient for calculating the temperature dependence of interface free energies \cite{tim,Frolov09jcp,tim_1,tim_2,Laird09,Laird2010}. 
    
      Before integrating Eq.~\eqref{eq:GH_steps} we need to choose $X$ and $Y$ since the selection of these variables determines which quantities need to be calculated to perform the thermodynamic integration. A convenient choice for the applications considered here is $X = A$ and $Y = N$. In this case Eq.~\eqref{eq:GH_steps} becomes
      \begin{equation}
        \label{eq:ti_3}
        \d{\fp{\gamma^\t{st} L}{T}} = - \Bigg( \f{[E]_{AN}}{T^2} - \sum_{i,j}^{x,y} \f{[\tau_{ij}]_{AN}L}{T} \td{\varepsilon_{ij}}{T} \Bigg) \d{T},
      \end{equation}
      where $[\tau_{ij}]_{AN}$ is given by Eq.~\eqref{eq:step_tau_NA}. Equation \eqref{eq:ti_3} can be integrated along a reversible thermodynamic path, where the temperature is increased from $T_0$ to $T$ while the solid is expanded to accommodate the thermal expansion, effectively maintaining zero bulk stress, \ie, $\gv{\sigma}^\t{b} = 0$. Notice that this thermodynamic path couples the \textit{a priori} independent variables $T$ and $\varepsilon$:
      \[
        \alpha_{ij} \equiv \l(\pd{\varepsilon_{ij}}{T}\r)_{\gv{\sigma}^\t{b} = 0} 
      \]
      where $\alpha_{ij}$ is the linear thermal-expansion factor and $i$ and $j$ are equal to $x,y, \t{ or } z$. We will assume here that the crystal lattice has cubic symmetry, allowing us to define our coordinate system in a way that eliminates the dependence of $\alpha_{ij}$ on the indexes $i$ and $j$. One further implication of following this thermodynamic path is that the system is not subject to shear strain during the thermal expansion; hence $[\tau_{ij}]_{AN}$ for $i \ne j$ performs no mechanical work. With these considerations Eq.~\eqref{eq:ti_3} becomes
      \begin{equation}
        \d{\fp{\gamma^\t{st} L}{T}} = - \Bigg( \f{[E]_{AN}}{T^2} - \f{2 \alpha \, [\tau_\t{avg}]_{AN}L}{T} \Bigg) \d{T}
        \label{eq:dgamma}
      \end{equation}
      where
      \begin{equation}
        [\tau_\t{avg}]_{AN} = \f{[\tau_{xx}]_{AN} + [\tau_{yy}]_{AN}}{2}
        \label{eq:tau_avg}
      \end{equation}
      is the average step stress. Upon integration of Eq.~\eqref{eq:dgamma} following the thermodynamic path described above we obtain
      \begin{align}
        \label{eq:TI_step}
        \gamma^\t{st}(T) = & \f{T}{T_0} \f{\gamma^\t{st}(T_0) \, L(T_0)}{L(T)} \\
                           & - \f{T}{L(T)} \int_{T_0}^T \Bigg( \f{[E]_{AN}}{T'^2} - \f{2 \alpha \, [\tau_\t{avg}]_{AN}L}{T'} \Bigg) \d{T'}. \nonumber
      \end{align}
      Note that all quantities inside the integral depend on the temperature $T'$.
    
      Equation \eqref{eq:TI_step} allows for the calculation of the temperature dependence of $\gamma^\t{st}$ if we know how to calculate all quantities on its right-hand side. The excess quantities inside the integral on the right-hand side of Eq.~\eqref{eq:TI_step}, $[E]_{AN}$ and $[\tau_\t{avg}]_{AN}$, can be computed readily from atomistic simulations since they only involve the calculation of the system energy and stress tensor. Thus, the only remaining term on the right-hand side of Eq.~\eqref{eq:TI_step} is the step excess free energy at a reference temperature $ \gamma^\t{st}(T_0)$. This type of term, present in all thermodynamic integration methods, cannot be trivially computed using atomistic simulations, since it involves the calculation of the absolute free energy of the system. In the next section we present a method due to Frenkel and Ladd \cite{frenkel_ladd, freitas_fl} which enables the calculation of the absolute free energy of solid systems. In the present context, this method enables the calculation of $\gamma^\t{st}(T_0)$ provided the temperature $T_0$ is chosen low enough such that the steps are structurally ordered (\ie, without an appreciable concentration of kinks, adatoms, or vacancies). Once we know the free energy of the step at this reference temperature, we can use Eq.~\eqref{eq:TI_step} to compute the absolute free energy of the step at any other temperature $T$ from values of $[E]_{AN}$ and $[\tau_\t{avg}]_{AN}$ at temperatures between $T_0$ and $T$.

    \subsection{\label{sec:theory_fl} Application of Frenkel-Ladd approach for calculation of step free energies}
      The Frenkel-Ladd \cite{frenkel_ladd} (FL) method is a type of thermodynamic integration approach that allows calculation of the absolute free energy of crystalline solids from atomistic simulations. Consider a system composed of $N$ identical particles with the Hamiltonian
      \begin{equation}
        \label{eq:H0}
        H_0 = \sum_{i=1}^N \f{\v{p}_i^2}{2m} + U(\v{r}_1, \v{r}_2, ..., \v{r}_N)
      \end{equation}
      where $m$ is the mass of the particles and $U(\v{r}_1, \v{r}_2, ..., \v{r}_N)$ is a many-body interatomic potential. We assume that, at the temperature and pressure of interest, the system's stable phase is a solid with a known crystalline lattice structure. Considering this lattice structure we will construct a second Hamiltonian for a reference Einstein crystal, which consists of particles of the same mass $m$ attached to the equilibrium lattice sites by harmonic springs with spring constant $k$:
      \begin{equation}
        \label{eq:HE}
        H_\t{E} = \sum_{i=1}^N \f{\v{p}_i^2}{2m} + \sum_{i=1}^N \f{1}{2} k \l( \v{r}_i - \v{r}_i^0 \r)^2
      \end{equation}
      where $\v{r}_i^0$ is the equilibrium lattice position of particle $i$ in the system described by $H_0$.
    
      In the FL method we use a Hamiltonian which is a linear interpolation of the Hamiltonians given by Eqs.~\eqref{eq:H0} and \eqref{eq:HE}:
      \begin{equation}
        \label{eq:H_interpolation}
        H(\lambda) = (1-\lambda) H_0 + \lambda H_\t{E},
      \end{equation}
      where $\lambda$ is a parameter of this Hamiltonian. The free energy of the system $H(\lambda)$ is
      \begin{equation}
        \label{eq:F_lambda}
        F(N, V, T; \lambda) = -\kB T \ln \l\{ \int \f{\d{\v{x}}}{h^{3N}} \exp \l[ -\beta H(\lambda) \r] \r\}
      \end{equation}
      where $\kB$ is the Boltzmann constant, $h$ is the Planck constant, $\v{x} = \{\v{r}_1, \v{r}_2, ..., \v{r}_N, \v{p}_1, \v{p}_2, ..., \v{p}_N\}$ is a point in the phase space of the particles of this system, and $\beta = 1 / \kB T$. It can be easily shown, by computing the derivative of Eq.~\eqref{eq:F_lambda}, that
      \[
        \pd{F}{\lambda} = \Avg{\pd{H}{\lambda}}_\lambda
      \]
      where $\avg{\ldots}_\lambda$ is the canonical ensemble average for a specific value of the parameter $\lambda$. From direct integration of the equation above from $\lambda = 0$ to $\lambda = 1$ we obtain
      \begin{equation}
        \label{eq:F0}
        F_0(N, V, T) = F_\t{E}(N, V, T) + \int_0^1 \avg{U-U_\t{E}}_\lambda \d{\lambda}
      \end{equation}
      where $F_0(N, V, T) \equiv F(N, V, T; \lambda = 0)$ is the free energy of the solid described by $H_0$, $F_\t{E}(N, V, T) \equiv F(N, V, T; \lambda = 1)$ is the free energy of the Einstein crystal, and $U_\mathrm{E}$ is the potential energy of the harmonic springs in the Einstein crystal. Since $H_\t{E}$ is composed of independent harmonic oscillators we can calculate its free energy analytically:
      \begin{equation}
        \label{eq:FE}
        F_\t{E}(N, V, T) = 3 N \kB T \ln \fp{\hbar \omega}{\kB T},
      \end{equation}
      where $\omega = \sqrt{k/m}$ is the natural frequency of the harmonic oscillators. 
    
      Equations \eqref{eq:F0} and \eqref{eq:FE} allow calculation of the absolute free energy of the solid $H_0$ from atomistic simulations. The only unknown in Eq.~\eqref{eq:F0} is the integrand on the right-hand side, which is an equilibrium ensemble average and, therefore, can be calculated directly using atomistic simulation techniques \cite{frenkel_smit} such as molecular dynamics or Monte Carlo with the Hamiltonian given by Eq.~\eqref{eq:H_interpolation}. The evaluation of Eq.~\eqref{eq:F0} can be performed in a straightforward manner using equilibrium simulations to obtain averages necessary to calculate the integral on the right-hand side numerically. However, this is an inefficient way to perform this thermodynamic integration. State-of-the-art methods \cite{freitas_fl} for evaluating Eq.~\eqref{eq:F0} based on nonequilibrium simulations have been developed and are now implemented in high-performance atomistic simulation software such as LAMMPS \cite{lammps} (Large-scale Atomic/Molecular Massively Parallel Simulator). These methods drastically reduce the computational cost of the thermodynamic-integration calculation and provide robust error-control criteria. An in-depth description of these techniques and detailed account of how they can be implemented in practice is given in Ref.~\onlinecite{freitas_fl}.
    
      We have shown in Sec.~\ref{sec:steps_theory}, Eqs.~\eqref{eq:gamma_excess} and \eqref{eq:gamma_st}, that the step free energy $\gamma^\t{st}(T_0)$ is a quantity that can be computed from the free energies of the three different regions shown in Fig.~\ref{fig:step_config}. Hence, our approach in the work presented below is to obtain the $\gamma^\t{st}(T_0)$ using the FL method to compute the absolute free energies of the relevant required systems. In so doing we have followed closely the methodology described in Ref.~\onlinecite{freitas_fl} to perform the FL calculations. Note, however, that the FL method has its applicability limited to low-temperature surfaces (flat and stepped), since at high temperatures the presence of surface vacancies, adatoms, and kinks on the steps breaks the FL method assumption that the atomic motion occurs around the equilibrium lattice positions, Eq.~\eqref{eq:HE}. Thus, the free energy computed with the FL method is only used as an initial integration point for the thermodynamic integral approach of Sec.~\ref{sec:TI_for_steps}, more specifically in Eq.~\eqref{eq:TI_step}.

  \section{\label{sec:simulations} Atomistic simulations}
    \subsection{Methodology}
      To demonstrate the application of the methodology described in the previous section, for computing step free energies by atomistic simulations, we focus on the $(111)$ surface of face-centered-cubic Cu, modeled with the embedded-atom-method (EAM) interatomic potential due to \citet{mishin_cu}. In previous simulations it has been found that this surface remains faceted at all temperatures up to the melting point of the EAM model ($T_\t{m} = 1327\K$ for the potential model considered \cite{tim_3}). No evidence for surface premelting was observed in these previous simulations, such that the surface maintains the layered crystalline structure up to $T_\t{m}$. Since the surface remains faceted, the step free energies are expected to remain finite up to this temperature.
    
      We have chosen molecular dynamics (MD) as the atomistic simulation technique to evaluate the step excess quantities necessary for the thermodynamic integration equations. All calculations were performed using LAMMPS \cite{lammps}, an open source implementation of MD. The Langevin thermostat \cite{langevin} was employed to sample the phase space, according to the canonical ensemble distribution. The relaxation time used for the thermostat was $\tau_\t{L} \equiv m / \gamma = 20 \t{ps}$, where $\gamma$ is the friction parameter and $m$ is the atomic mass. The timestep was chosen based on the highest-frequency normal mode of the system ($\nu_\t{max} = 7.8 \times 10^{12} \t{ Hz}$); we have taken $\Delta t$ to be approximately $1/60$th of the oscillation period of that normal mode: $\Delta t = 2 \,\t{fs}$.
    
    \subsection{System geometry and dimensions \label{sec:simulations_dimensions}}
      In Sec.~\ref{sec:TI_theory} Eq.~\eqref{eq:TI_step} was derived for the temperature dependence of $\gamma^\t{st}$, based on the choice $X = A$ and $Y = N$. In this sub-section we elaborate further why this is a convenient choice for the calculation of the excess quantities that appear in Eq.~\eqref{eq:TI_step} from atomistic simulations. From Eq.~\eqref{eq:excess_definition} we have
      \begin{equation}
        [Z]_{AN} = \f{\Dmat{Z^\t{st}}{A^\t{st}}{N^\t{st}}{Z^\t{t}}{A^\t{t}}{N^\t{t}}{Z^\t{b}}{A^\t{b}}{N^\t{b}}}{\dmat{A^\t{t}}{N^\t{t}}{A^\t{b}}{N^\t{b}}}.
        \label{eq:excess_choice_1}
      \end{equation}
      The bulk region in Fig.~\ref{fig:step_config} does not have any surface which means $A^\t{b} = 0$. Heretofore the regions shown in Fig.~\ref{fig:step_config} had arbitrary dimensions; from now on we choose the dimensions of the step and terrace regions in such a way that they have the same surface area, \ie, $A^\t{st} = A^\t{t}$. Furthermore, we choose the depth of these regions such that they contain the same number of atoms: $N^\t{st} = N^\t{t}$. With this particular choice of dimensions, the excess quantities shown in Eq.~\eqref{eq:excess_choice_1} become $[Z]_{AN} = Z^\t{st} - Z^\t{t}$. Thus, the need to compute the thermodynamic properties for the bulk ($Z^\t{b}$) is minimize and $[Z]_{AN}$ becomes a simple difference between the properties of the step and terrace regions.
    
\begin{figure}
        \includegraphics[width=0.48\textwidth]{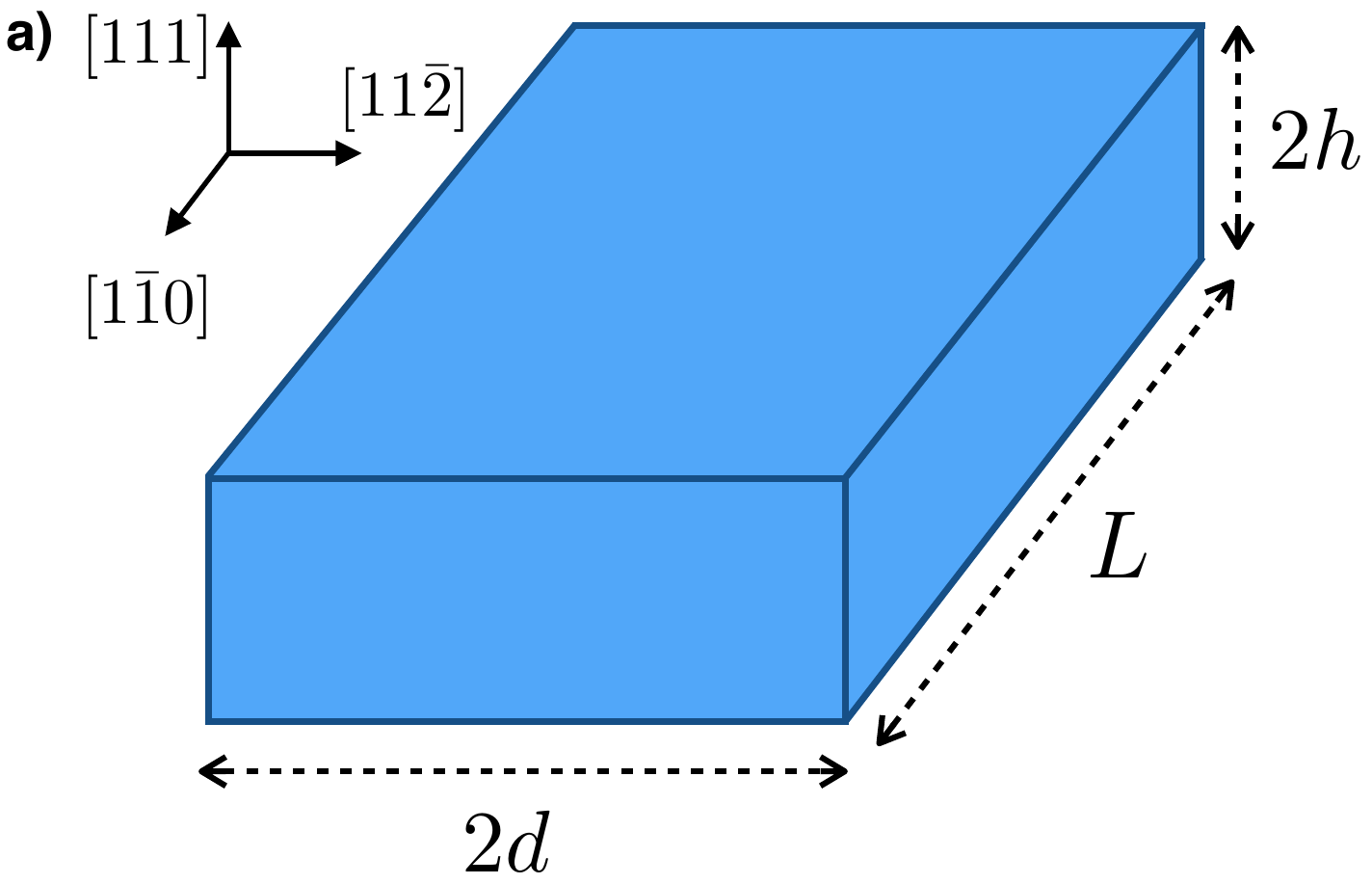}
        \vspace{0.5cm}
        \includegraphics[width=0.48\textwidth]{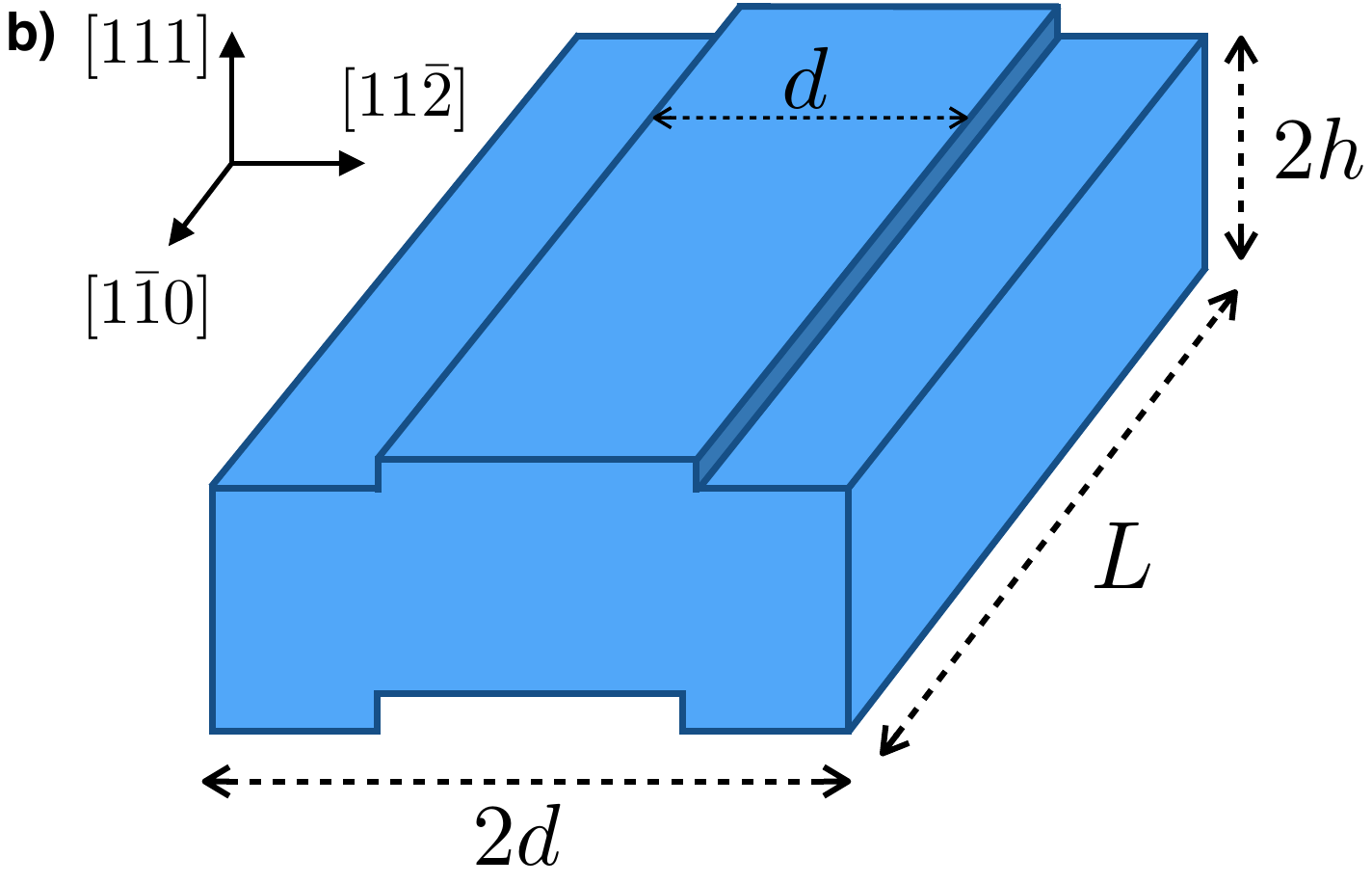}
        \caption{\label{fig:geometry_box} Simulation box geometry for (a) the system with a flat terrace and (b) the system containing steps. $d$ denotes the lateral step separation distance, $h$ the bulk depth, and $L$ the step length. The step line is parallel to the $[1\bar{1}0]$ direction and on the $(111)$ surface plane. The step separation, $d$, is measured along the $[11\bar{2}]$ direction, perpendicular to the step line direction. Periodic boundary conditions are applied along the $[1\bar{1}0]$ and $[11\bar{2}]$ directions, within the surface plane. The systems illustrated in (a) and (b) are constructed such that they contain the same number of atoms and have the same total surface area.}
      \end{figure}

      To calculate step excess quantities we modeled two different simulation blocks illustrated in Fig.~\ref{fig:geometry_box}. The simulation block shown in Fig.~\ref{fig:geometry_box}a is a solid film with two flat $(111)$ surfaces. Periodic boundary conditions were applied for the directions parallel to the surface. The second simulation block illustrated in Fig.~\ref{fig:geometry_box}b was obtained from the first one by adding half of an atomic plane on the top surface and removing half of the atomic plane from the bottom surface. As a result of this construction, the second block has four surface steps. At the same time the construction ensures that the two simulation blocks have the same number of atoms and the same surface area. Properties $Z^\t{st}$ and $Z^\t{t}$ were then calculated for the two blocks with and without steps, respectively. The difference between these quantities gives the step excess $[Z]_{AN}$ given by Eq.~\eqref{eq:excess_choice_1}. Indeed, $Z^\t{st}-Z^\t{t}$ represents the excess of property $Z$ due to steps, when the reference system has the same surface area and the same number of atoms. We remind the reader that the excess quantities inside the integral on the right-hand side of Eq.~\eqref{eq:TI_step} are $[E]_{AN} = E^\t{st} - E^\t{t}$ and $[\tau_{ii}]_{AN} = (\sigma_{ii}^\t{st} V^\t{st} - \sigma_{ii}^\t{t} V^\t{t})/L$ and can be readily computed from atomistic simulations since they involve only the calculation of the energy and stress tensor of each of the systems in Fig.~\ref{fig:geometry_box}. 
    
\begin{figure}
        \includegraphics[width=0.48\textwidth]{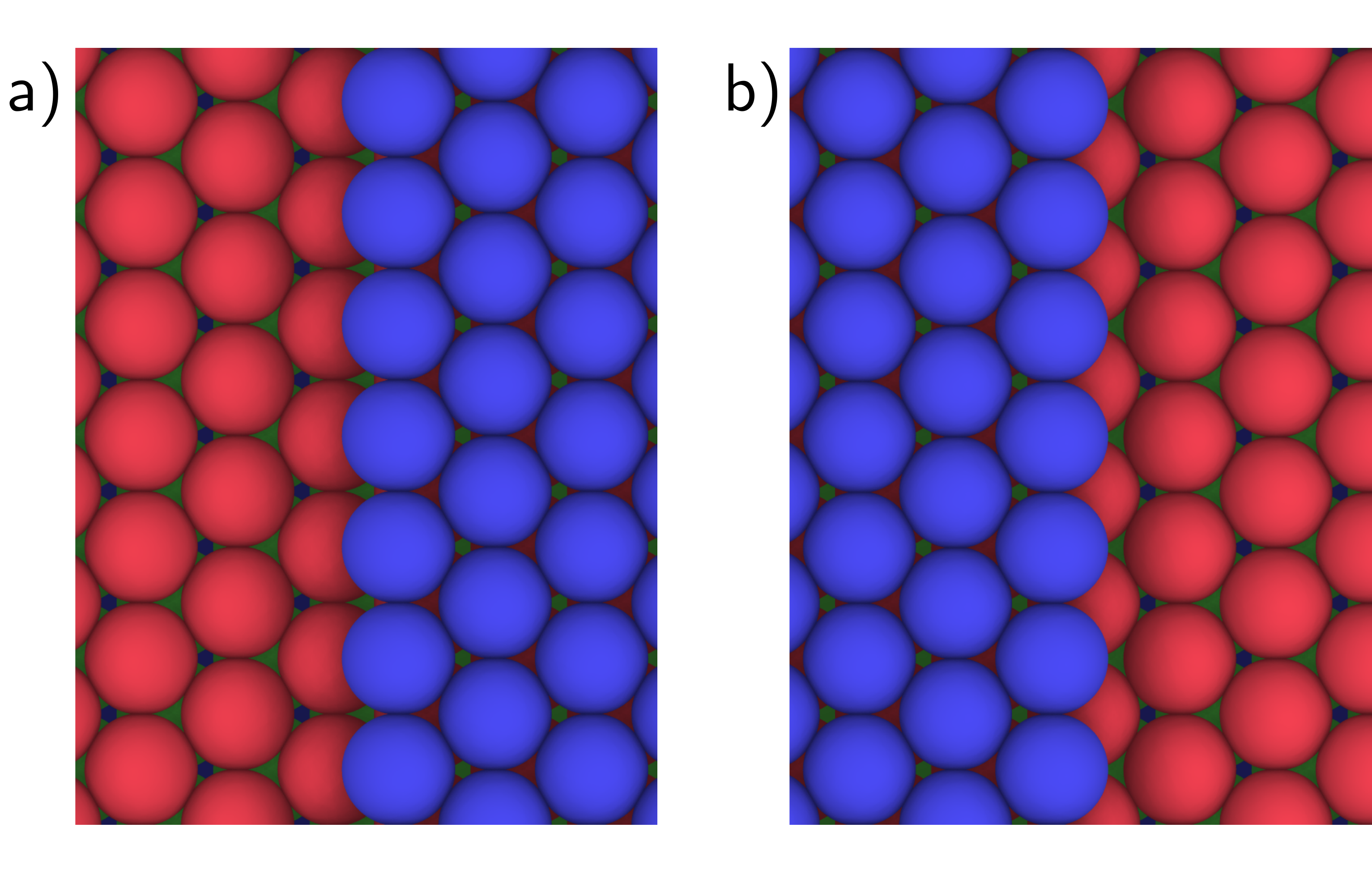}
        \caption{\label{fig:geometry_step} Atomic configuration of (a) $\avg{110}$A and (b) $\avg{110}$B steps on the $(111)$ surface of an fcc lattice. Atoms are colored according to the atomic layer they belong to: blue atoms belong to the first layer, red atoms to the second layer, and yellow atoms to the third layer. Atoms at the edge of the two different $\avg{110}$A and $\avg{110}$B steps have different nearest-neighbor configurations on the $(111)$ plane immediately below the surface.}
      \end{figure}

      The steps considered in the MD simulations of the simulation cells illustrated by Fig.~\ref{fig:geometry_box}b are directed along the close-packed $\avg{110}$ direction. The crystallographic symmetry of the $(111)$ surface is such that the two steps shown in each of the surfaces of Fig.~\ref{fig:geometry_box}b are slightly different; it can be seen in Fig.~\ref{fig:geometry_step} that they have different nearest-neighbor configurations on the $(111)$ plane immediately below the surface. The step with lowest zero-temperature energy \cite{muller} ($U_0 = 103.13 \, \t{meV}/\Ang$) is a $\avg{110}$A step, while the step with the slightly larger energy ($U_0 = 104.08 \, \t{meV}/\Ang$) is a $\avg{110}$B step. Using the terminology of Ref.~\onlinecite{dft_1} $\avg{110}$A steps have $\avg{100}$ microfacets and $\avg{110}$B steps have $\avg{111}$ microfacets.
    
      We have chosen the simulation box size in such a way that the step-step interaction energy of all four steps in Fig.~\ref{fig:geometry_box}b was negligible compared to the step self-energy (\ie, the energy of an isolated step). Steps are abrupt interruptions of the surface first layer; hence, they deform the atomic structure around them, creating an elastic field \cite{marchenko}. The interaction energy due to the overlap of the strain fields of the two steps decays as $d^{-2}$ with the step-step separation and exponentially with the bulk depth (see Ref.~\onlinecite{srolovitz} and references therein). Following the work of \citet*{srolovitz} we have verified this behavior for the step elastic interaction energy \footnote{If these interactions are appreciable we would expect a significant difference between results obtained from supercells with symmetric (as shown in Fig.~\ref{fig:geometry_box}) versus antisymmetric indentations on the top and bottom surfaces. But if the crystal slab is thick enough to minimize step interactions the differences would be negligible. We performed a separate independent test of the effect of system size, by comparing the step self energy derived from supercells with the geometry shown in Fig.~\ref{fig:geometry_box}, with results obtained using non-orthogonal boundary conditions \cite{srolovitz} where there is only one step on each surface. The values for the step self energy agreed for both geometries to within $0.01\%$.} of our model and we have determined the step-step distance ($d$) and bulk depth ($h$) such that $E_\t{int}/U_0 \le 10^{-4}$, where $E_\t{int}$ is the total step interaction energy and $U_0$ is the step self-energy at zero temperature. The box dimensions obtained are $d = 70.8\Ang$ and $h = 53.2\Ang$ at $T = 0\K$; for finite temperatures we have increased the system dimensions to account for thermal expansion, for zero bulk stress.
    
      The simulation box length along the step line cannot be determined based on static simulations. In order to determine the step length necessary to eliminate finite-size effects along the step direction it is necessary to consider fluctuations of the step line that appear at finite temperatures, known as capillary fluctuations. The accurate evaluation of step excess quantities requires satisfactorily sampling the normal modes of these fluctuations (\ie, the capillary waves) during the simulation. If the step length used is too small the sampling of long-wavelength modes is suppressed. On the other hand, an excessively lengthy step would make the thermodynamic integration calculations prohibitively long due to the need to sample normal modes with very long wavelengths and associated long relaxation times. Thus, to determine the step length required in the simulations we need to analyze the convergence of $[E]_{AN}$ and $[\tau_\t{avg}]_{AN}$ with the step length since, according to Eq.~\eqref{eq:TI_step}, the thermodynamic integration equation depends on the computation of these two quantities.
      
\begin{figure}
        \includegraphics[width=0.48\textwidth]{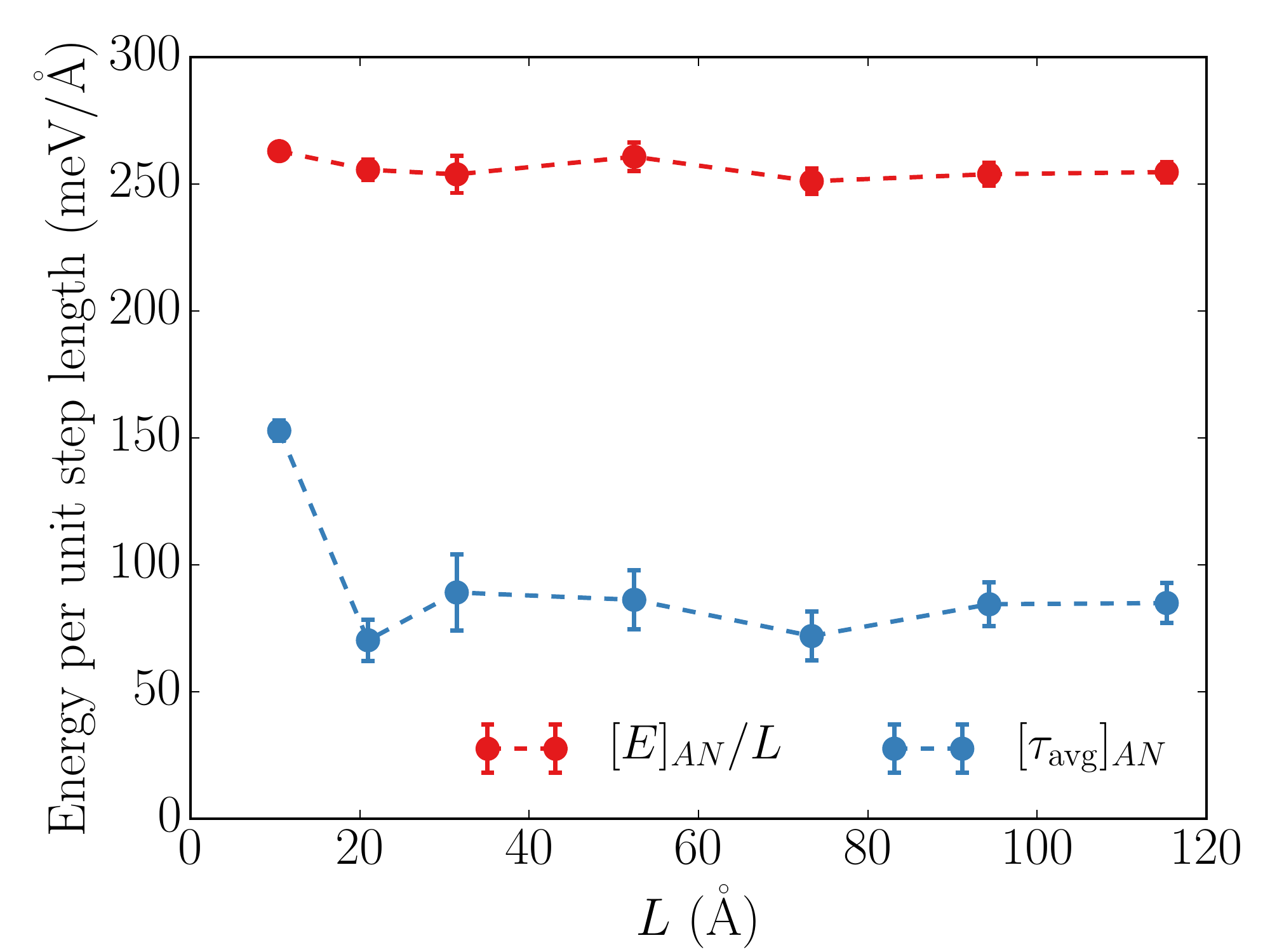}
        \caption{\label{fig:excess_vs_L}Convergence of step excess energy ($[E]_{AN}$) and average step excess stress ($[\tau_\t{avg}]_{AN}$) with step length ($L$) at $T = 1300\K$. Error bars are the standard error of the mean, computed by taking into consideration the relaxation times of the capillary wave normal modes.}
      \end{figure}

      Using the values of $d$ and $h$ determined above we have run simulations at $T = 1300\K$ for systems with different step lengths ($L$) and calculated the step excess energy and stress. Figure~\ref{fig:excess_vs_L} shows the convergence of the step excess quantities with step length for these simulations. Based on these results we have chosen $L = 30.7 \Ang$ as the step length for the next simulations since the step excess quantities are seen to be well converged for steps of this size.

  \section{\label{sec:results} Results and discussion}
    \subsection{Step free energies from Frenkel-Ladd simulations}
\begin{figure}
        \centering
        \includegraphics[width=0.48\textwidth]{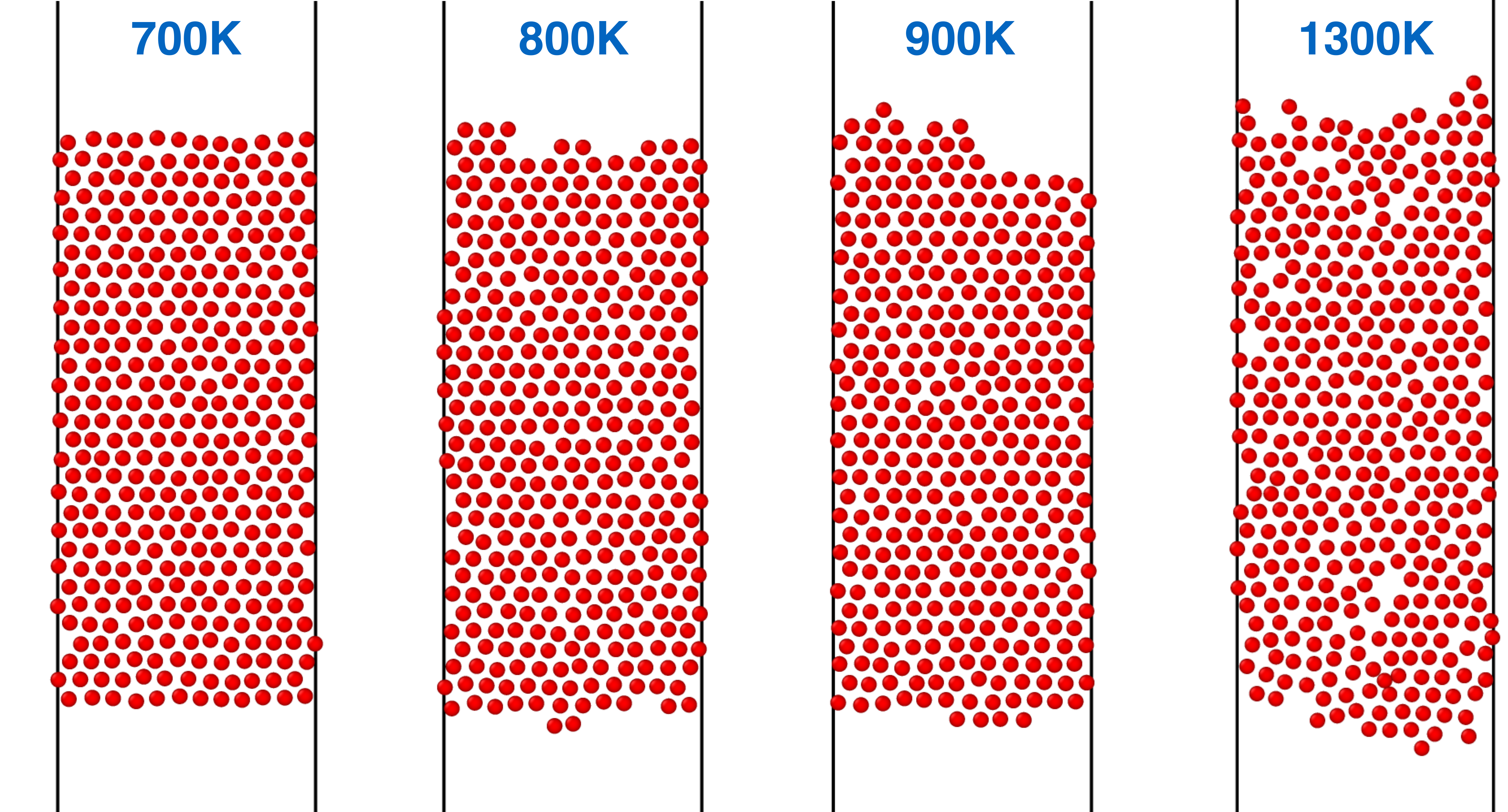}
        \caption{\label{fig:step_profile}Plan view of two steps on the $(111)$ surface, showing only atoms on the top step layer. Variations in the step position due to capillary fluctuations become larger as we raise the temperature and approach the melting point.}
      \end{figure}

      Using the box dimensions specified in Sec.~\ref{sec:simulations_dimensions}, we have constructed two systems to be used in the FL simulations, one with a flat surface and the another with stepped surfaces, as shown in Figs.~\ref{fig:geometry_box}a and b. Both systems have $39 168$ atoms and the same surface area. In Fig.~\ref{fig:step_profile} we show plan-view snapshots \cite{ovito} of the top layer of atoms from typical configurations for the stepped system at different temperatures. At temperatures of $700\K$ or lower, the step line is mostly straight with small fluctuations due to atomic vibrations, while as we raise the temperature closer to the melting point ($T_\t{m} = 1327\K$) configurational disorder due to capillary fluctuations and the formation of vacancies and adatoms becomes pronounced. The presence of appreciable configurational disorder limits the application of the FL method to temperatures below $700\K$. The formation of defects above this temperature causes a sharp increase in the dissipation during the switching to the Einstein crystal. This excessive dissipation is characteristic of irreversible processes and violates the assumptions necessary for the derivation of the FL method equations, namely, the reversibility of the integration path and that the atomic motion occurs around average positions given by the equilibrium lattice positions.
    
      The FL method was applied to both systems in Fig.~\ref{fig:geometry_box} according to the nonequilibrium techniques presented in Ref.~\onlinecite{freitas_fl}. We have employed a switching time of $t_\t{s} = 4\ns$ and the S-shaped \cite{freitas_fl,dekoning} functional form for the $\lambda(t)$ parameter. The simulations were carried out at temperatures ranging from $100\K$ to $700\K$ in intervals of $100\K$. Estimates for the statistical errors were obtained by performing three independent switching simulations (forward and backward) for each temperature.
    
\begin{figure}
        \includegraphics[width=0.48\textwidth]{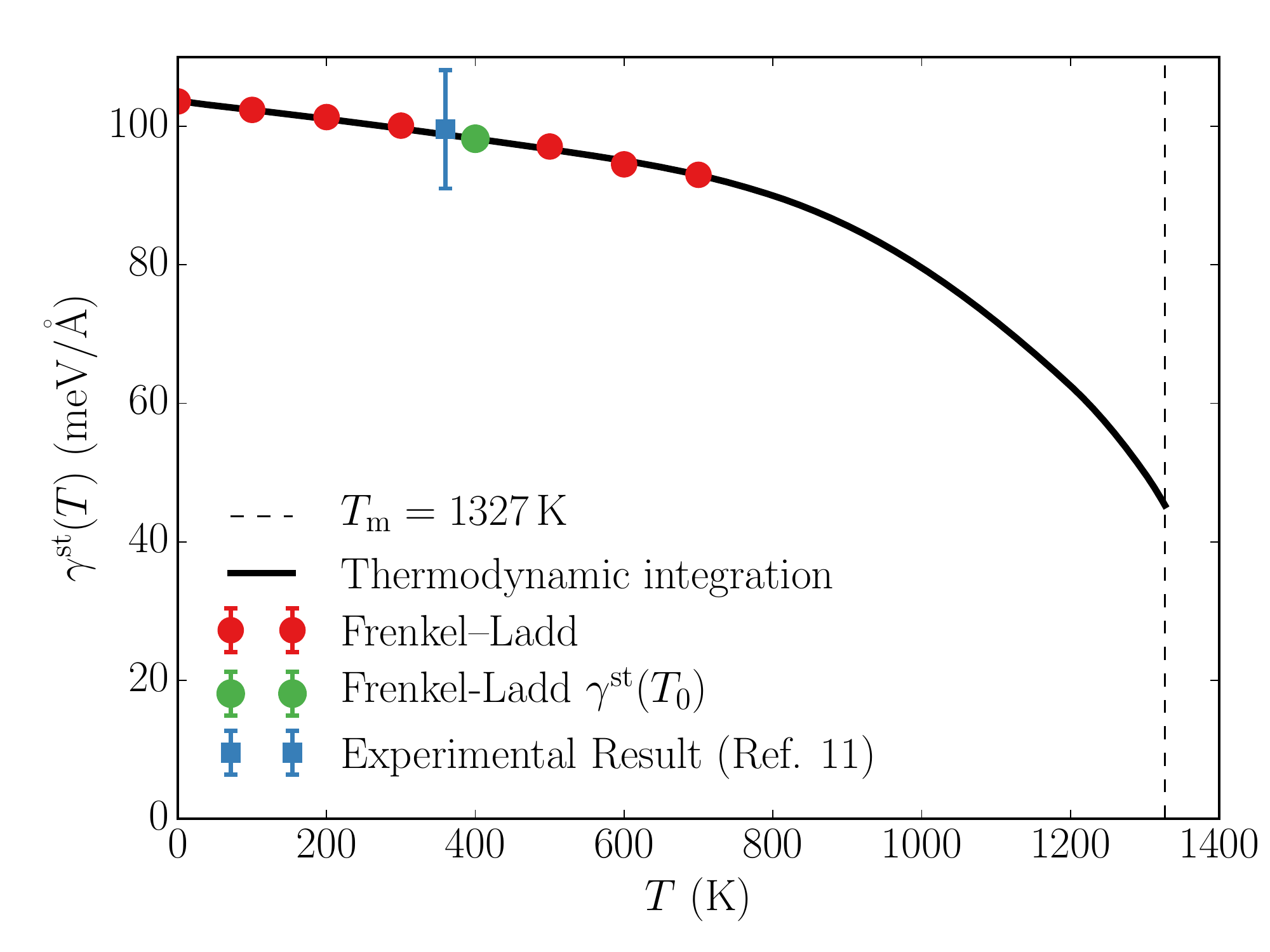}
        \caption{\label{fig:gamma_step}Temperature dependence of the calculated average of $\avg{110}$A and $\avg{110}$B step free energies on the (111) surface of elemental Cu. The black solid line was calculated from the step excess quantities using the thermodynamic integration method described in Sec.~\ref{sec:TI_for_steps}. The initial point for the integration $\gamma^\t{st}(T_0)$ was obtained using the Frenkel-Ladd method at $T_0 = 400\K$. Extra calculations using the Frenkel-Ladd method were performed for $T \ne T_0$; they are shown as red dots in the figure and are in excellent agreement with the independent thermodynamic integration results. Also included in the figure is an experimental measurement of the step free energy at $T = 360 \K$, taken from Ref.~\onlinecite{exp_4}, which is seen to be very close in magnitude to the computed value at this temperature.}
      \end{figure}

      Based on the discussion in Sec.~\ref{sec:simulations_dimensions}, the step excess free energy was calculated from the difference of the free energy of the two systems in Fig.~\ref{fig:geometry_box}: $\gamma^\t{st} = (F^\t{st} - F^\t{t}) / L$, where $L$ is the total length of the four steps in Fig.~\ref{fig:geometry_box}b. Since the surfaces in the system illustrated in Fig.~\ref{fig:geometry_box}b contain both $\avg{110}$A and $\avg{110}$B types of steps, the FL method provides the average of the free energy of both of these step types. The results of the FL simulations are shown as the red and green dots in Fig.~\ref{fig:gamma_step}. Note that the error bars are smaller than the points on the plot. The standard error of the mean of the points in Fig.~\ref{fig:gamma_step} is $\approx 1 \, \t{meV}/\Ang$, which requires the calculation of the free energy per atom for each system, which is achieved with an accuracy of $\approx3 \, \mu\t{eV/atom}$. Such high statistical accuracy is achievable due to the high efficiency and accuracy of the nonequilibrium Frenkel-Ladd method used in this work. Further details about the technique as well as an in-depth analysis of error control and estimation are provided in Ref.~\onlinecite{freitas_fl}.

    \subsection{Step excess quantities}
\begin{figure}
        \includegraphics[width=0.48\textwidth]{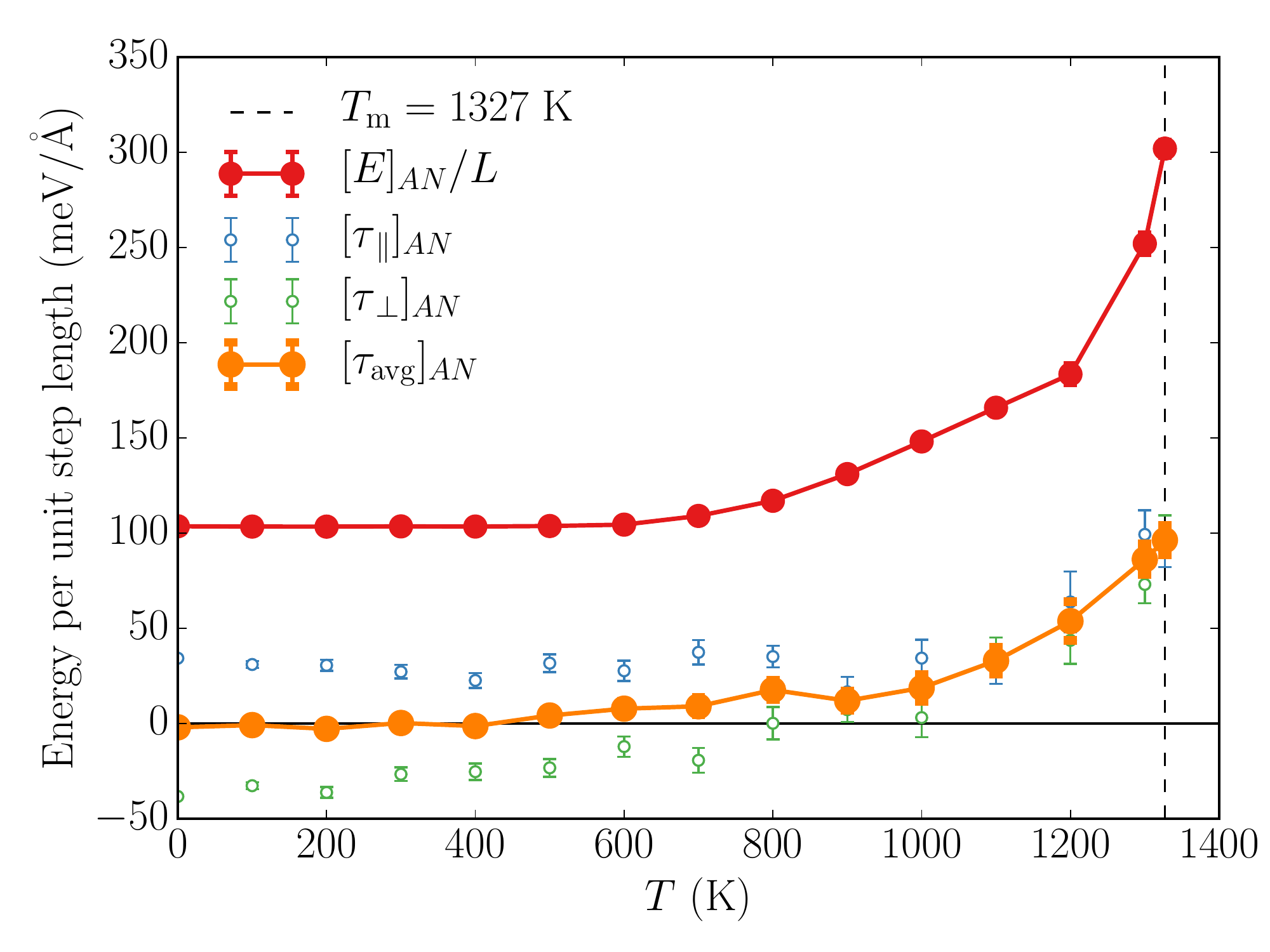}
        \caption{\label{fig:excess_vs_T}Temperature dependence of step excess energy $[E]_{AN}$ and step excess stress $[\tau_\t{avg}]_{AN}$. $[\tau_\t{avg}]_{AN}$ is the average of the step stresses parallel ($[\tau_\parallel]_{AN}$) and perpendicular ($[\tau_\perp]_{AN}$) to the step line. Notice how the stress perpendicular to the step line changes from tensile to compressive as the temperature increases. Error bars corresponding to the standard error of the mean were obtained for all data points, although they are smaller than the symbols employed for some of the data points.}
      \end{figure}

      The step excess quantities were calculated for systems with the same size and number of atoms as the systems used for the FL calculations. From the MD simulations we obtained the average energy of the systems with the step $E^\t{st}$ and the flat terrace $E^\t{t}$, and also the components of stress tensor $\gv{\sigma}^\t{st}$ and $\gv{\sigma}^\t{t}$. The step excess properties $[E]_{AN}$ and $[\tau_\t{avg}]_{AN}$ were then computed by taking the difference between the quantities of the stepped system and the flat-terrace system, as described in Sec.~\ref{sec:simulations_dimensions}. The MD simulations were performed for temperatures ranging from $100$ to $1300\K$, in intervals of $100\K$. Additionally, we also performed one simulation at the melting temperature for the potential $T_\t{m} = 1327\K$. The systems were equilibrated for $6\ns$ before calculating the values of $E^\t{st}$, $E^\t{t}$, $\gv{\sigma}^\t{st}$, and $\gv{\sigma}^\t{t}$. After equilibration, these values were sampled at intervals of $2\,\t{ps}$ for $400\ns$ at each temperature. Figure \ref{fig:excess_vs_T} shows the temperature dependence of $[E]_{AN}$ and $[\tau_\t{avg}]_{AN}$. The error bars correspond to the standard error of the mean for each data point, obtained through a block average analysis of the data collected for each temperature. Note that the error bars of $[E]_{AN}$ are too small to appear on the plot.
      
      The results for $[E]_{AN}$ in Fig.~\ref{fig:excess_vs_T} show that the excess energy increases with temperature from $(103.61 \, \pm \, 0.02) \, \t{meV/}\Ang$ at $T=0\K$ to $(302\, \pm \, 4) \, \t{meV}/\Ang$ at $T_\t{m}$. Within a large temperature interval, from zero to approximately $800 \K$ (\ie, a homologous temperature of approximately $0.60$), the value of $[E]_{AN}$ remains essentially constant, and then begins to increase much more rapidly as the melting temperature is approached. The simulations show that $[E]_{AN}$ remains finite, and does not diverge as the melting point is approached.
    
      The results for excess stress in Fig.~\ref{fig:excess_vs_T} show that $[\gv{\tau}]_{AN}$ is appreciably anisotropic at low $T$: the step stress component perpendicular to the step, $[\tau_\perp]_{AN}$, is compressive at low temperatures while $[\tau_\parallel]_{AN}$ is tensile. Although they are similar in magnitude we notice that they have measurably different values at $0\K$: $[\tau_\perp]_{AN} = - 38.3 \, \t{meV/}\Ang$ and $[\tau_\parallel]_{AN} = 34.3 \, \t{meV/}\Ang$. Both parallel and perpendicular components increase with temperature, but the perpendicular does so faster, the consequence being that the anisotropy becomes reduced at high temperatures, where both components become compressive. Notice also that the excess average stress, $[\tau_\t{avg}]_{AN}$, remains almost constant for low temperatures, before the onset of large capillary fluctuations of the step. As for $[E]_{AN}$, only for temperatures above approximately $800 \K$ is a significant temperature dependence of the step excess stress observed.
    
    \subsection{Step free energies from thermodynamic integration calculations \label{sec:results_TI}}
      In this subsection we focus on the temperature-dependent step free energies, obtained by the thermodynamic-integration (TI) approach described in Sec.~\ref{sec:TI_for_steps}. The results obtained from this approach are shown as the solid line in Fig.~\ref{fig:gamma_step}, which plots the value of $\gamma^\t{st}$ over the entire temperature range from $T = 0\K$ up to $T_\t{m}$. In performing the TI calculations, we have chosen $T_0 = 400\K$ as the reference point for the thermodynamic integration, and the integration was performed in both directions, from $T_0$ to $T_\t{m}$ and from $T_0$ to $\approx 0\K$. As noted above, the TI values for $\gamma^\t{st}$ agree well with those from the FL method that were not used in the integration (red points in Fig.~\ref{fig:gamma_step}), demonstrating the consistency of the predictions for the temperature dependence of $\gamma^\t{st}$ at low homologous temperatures obtained from these two independent methods. We present in the \hyperref[sec:appendix]{Appendix} a discussion of the numerical convergence of the TI results, including error calculations and the independence of the final results on the choice of the reference temperature $T_0$.
    
      Overall, the TI results in Fig.~\ref{fig:gamma_step} show that the temperature dependence of $\gamma^\t{st}$ is large and highly nonlinear over the full temperature range. Although the magnitude of $\gamma^\t{st}$ remains finite at the melting point, indicating that the surface remains faceted up to $T_\t{m}$, the net effect of increasing temperature is a sizable decrease of $\gamma^\t{st}$. Specifically, increasing the temperature up to melting leads to a decrease in magnitude of $\gamma^\t{st}$ by more than half, from a value of $(103.61 \pm 0.02) \, \t{meV/}\Ang$ at $T=0\K$ to $(45.8 \pm 0.4) \, \t{meV/}\Ang$ at $T = T_\t{m}$.
    
      Considering the temperature dependence of $\gamma^\t{st}$ in further detail, we divide the results into two temperature ranges: low homologous temperature up to $800\K$ (\ie, from homologous temperatures of zero to approximately 0.60), and high homologous temperatures from $800\K$ up to the melting point. In the first temperature range, the excess quantities presented in the previous section are approximately constant in value, and $\gamma^\t{st}$ displays a relatively weak rate of decrease with temperature. Over this temperature range the value of $\gamma^\t{st}$ decreases approximately linearly, by roughly $13\%$ percent, from a value of $(103.61 \pm 0.02) \, \t{meV/}\Ang$ to $(90.3 \pm 0.2) \, \t{meV/}\Ang$. Since the steps are observed to remain straight on the simulation length and time scales (\ie., no evidence of appreciable kinks, adatoms, or surface vacancies is observed) for temperatures up to $800\K$, we interpret the temperature dependence of $\gamma^\t{st}$ over this temperature range to arise primarily from atomic vibrational contributions to the step excess thermodynamic quantities.
    
      Above $T=800\K$, $\gamma^\t{st}$ displays a much more pronounced temperature dependence. From $800\K$ up to $T_\t{m}$ the value of $\gamma^\t{st}$ decreases by roughly $51\%$ percent, from a value of $(90.3 \pm 0.2) \, \t{meV/}\Ang$ to $(45.8 \pm 0.4) \, \t{meV/}\Ang$. In this temperature range, the concentration of surface adatoms and vacancies increases significantly, and the magnitudes of the step capillary fluctuations become more pronounced. The larger temperature dependence of $\gamma^\t{st}$ over this temperature range is thus interpreted to be a manifestation of the effect of such configurational disorder on the step excess thermodynamic quantities.

    \subsection{Comparison with previous measured and calculated results}
      Although we are not aware of previous results presenting the temperature dependence of step free energies in Cu all the way up to the melting point, there have been measurements and previously published calculations at low temperatures for this system, to which the present simulation results can be compared.
    
      The step free energy of Cu$(111)$ $[110]$A and B steps at selected temperatures has been obtained experimentally from the analysis of adatom and vacancy islands observed using scanning tunneling microscopy \cite{exp_1, exp_2, exp_3, exp_4}. For a comprehensive comparison of two available methods for computing $\gamma^\t{st}$ experimentally, we refer the reader to Ref.~\onlinecite{exp_4}, where Steimer \textit{et al.} report $\gamma^\t{st} = 256 \, \pm \, 22 \, \t{meV/a}$ for an average of $A$ and $B$ steps, where $a$ is the atomic distance along the $[110]$ direction and the measurement is for an average temperature of $\overline{T} = 360\K$ ($T \in [280, 440] \K$). The present results are remarkably close to this value, as indicated in Fig.~\ref{fig:gamma_step}: we obtain values of $\gamma^\t{st} = 254.8 \, \pm \, 0.2 \, \t{meV/a}$ for the same temperature. Moreover, the temperature dependence of $\gamma^\t{st}$ shown in Fig.~\ref{fig:gamma_step} is consistent with the analysis in Ref.~\onlinecite{exp_4}, suggesting that the step free energy has a weak temperature dependence for the temperature range at which the experiments were conducted.
    
      This good agreement between the present simulation results and experimental measurements is achieved despite the approximations inherent in the classical description of the interatomic interactions by an EAM potential model. Importantly, a similar level of agreement is also obtained by the EAM model and available \textit{ab initio} values at zero temperature obtained by density functional theory (DFT). Specifically, the value of the step energy given by the EAM potential considered in this work is $\gamma^\t{st}=264.8 \, \t{meV/a}$ at $T=0\K$, which agrees well with the DFT result of $\gamma^\t{st} = 270 \, \t{meV/a}$ reported in Refs.~\onlinecite{dft_1,dft_2,dft_3}. The fact that the current results agree well with DFT at zero temperature, and with experiment at finite temperatures, suggests that the latter agreement is not a result of cancellation of errors resulting from inaccurate energetics and temperature dependencies. Rather the EAM model for Cu of Mishin {\em et al.} \cite{mishin_cu} employed in this work appears to yield accurate values for $\gamma^\t{st}$ and its temperature dependence, at least at low homologous temperatures.

  \section{\label{sec:summary} Summary}
    We present a thermodynamic formalism for steps on faceted surfaces of single-component crystalline solids, resulting in the derivation of a general adsorption equation, Eq.~\eqref{eq:adsorption}, relating changes in step free energy ($\gamma^\t{st}$) to variations in chemical potential, surface free energy, temperature, and strain. The rate of change of $\gamma^\t{st}$ with respect to variations in these variables is related to surface excess quantities of particle number, surface area, entropy, and stress, respectively. Due to the existence of Gibbs-Duhem relations for the bulk and surface, which give rise to constraints on the variations of the intensive variables, Cramer's rule can be used to express the adsorption equation in terms of a particular choice for the set of independent variables. The approach results in the definition of step excess quantities formulated in terms of determinants, following the formalism first introduced in the context of interfacial thermodynamics by Cahn \cite{cahn}. A direct result of the formulation developed in the present work is the definition of a step excess stress, Eq.~\eqref{eq:Shuttleworth_step_2}, which is the step analog of the familiar surface stress quantity, and which represents the excess force on the perimeter of a stepped surface due to the presence of a step. Although the formalism presented in this work is developed only for the special case of single-component crystalline surfaces, the underlying approach is more general, and can be extended to multicomponent/multiphase situations, as demonstrated recently by Frolov and Mishin \cite{tim2015}.

    The thermodynamic formalism presented in this work is demonstrated to provide a convenient framework for thermodynamic-integration calculations of the temperature dependence of $\gamma^\t{st}$ by atomistic simulations. For this purpose, it is natural to employ a particular choice for the set of independent intensive variables that leads to the definition of step excess quantities, in a manner that is similar to choosing a Gibbs \cite{gibbs_2} dividing surface leading to zero excess volume and particle number. By combining the resulting expression for the adsorption equation with the Gibbs-Helmholtz relation, we derive an expression for the temperature dependence of the step free energy, Eq.~\eqref{eq:dgamma}, in terms of step excess energy and excess stress quantities that can be readily calculated in atomistic simulations. It is straightforward to extend the proposed TI approach to steps at faceted solid-liquid interfaces, grain boundaries and phase boundaries in multicomponent systems \cite{tim2015}. This approach can provide full temperature and composition dependence of step free energy from atomistic simulations, provided that a reference free energy value is known at some temperature and composition. In the present work we have demonstrated how the Frenkel-Ladd method can be employed for this purpose, when the interfaces of interest involve only solid phases. For solid-liquid or solid-vapor interfaces alternative approaches would be needed such as those based on nucleation simulations (\eg, Ref.~\onlinecite{tim_jcp}) or analyses of capillary fluctuations (\eg, Refs.~\onlinecite{cwm_prl,solidsolid}).

    We demonstrate the application of the thermodynamic integration formalism for the case of $\avg{110}$ steps on faceted $\{111\}$ surfaces of element Cu, employing MD simulations based on a classical EAM potential due to Mishin \textit{et al.} \cite{mishin_cu}. By combining the thermodynamic-integration formalism with the the Frenkel-Ladd method for computing a reference value of $\gamma^\t{st}$ at low temperatures, where the step structure remains highly ordered, we present a calculation of the step free energy over the entire temperature range from zero up to the melting point.

    In the process of performing the thermodynamic-integration calculations, we compute temperature-dependent values for the step excess energies and stresses, as shown in Fig.~\ref{fig:excess_vs_T}. The excess energy is found to display a weak temperature dependence up to a homologous temperature of approximately 0.60; beyond this temperature the excess energy increases strongly as the step displays growing configurational disorder due to the formation of surface adatoms and vacancies and appreciable capillary fluctuations. For the step excess stress, we have obtained negative $[\tau_\perp]_{AN}$ and positive $[\tau_\parallel]_{AN}$ at low homologous temperatures, with both terms having similar magnitudes. With increasing temperature, the low-temperature anisotropy of the step stress is greatly reduced, and at high temperatures $[\tau_\perp]_{AN}$ becomes positive. Therefore, thermal effects such as thermal expansion, vibrational fluctuations, and configurational disordering affect each step stress component differently. It is worth noting that this behavior is not unique to step stresses; it has been observed before in atomistic simulations \cite{gamma,msmse} that the surface stress for solid-liquid interfaces also presents positive and negative values, depending on the system properties and thermodynamic conditions.
    
    For the temperature dependence of the calculated step free energy, our findings are shown in Fig.~\ref{fig:gamma_step} and can be summarized as follows. At low homologous temperatures (\ie, less than approximately 0.6), where the thermal effects are interpreted to be associated primarily with atomic vibrations, $\gamma_\t{st}$ is calculated to display a relatively weak temperature dependence. At these low temperatures, the calculated magnitudes of $\gamma^\t{st}$ show good agreement with previously reported experimental measurements and DFT calculations, indicating the accuracy of the employed EAM potential for the present application. The calculated temperature dependence of $\gamma^\t{st}$ increases strongly at higher homologous temperatures, as the step becomes increasingly configurationally disordered. The net effect is a reduction in the step free energy by more than half as the temperature is increased from zero up to the melting temperature. Such a strong temperature dependence at high homologous temperatures would be expected to have important consequences for kinetic processes such as surface island nucleation and growth kinetics.

    We emphasize that the formalism presented in this work provides a general framework for the calculation of step free energies for elemental systems using atomistic simulation methods, and it is applicable beyond the application demonstrated in this work for elemental Cu modeled by an EAM classical potential. In practical applications to other systems, several considerations should be taken into account. First, the Frenkel-Ladd approach provides a methodology to compute step free energies only at temperatures where contributions of configurational disorder due to kinks, adatoms, and vacancies can be ignored, \ie, where vibrational contributions to the temperature dependence of the excess properties dominate; to ensure that this is the case sufficiently long simulations are required to guarantee structural equilibration, or theoretical analyses based on calculated kink and point-defect formation energies should be performed. For temperatures where the steps remain structurally ordered, the Frenkel-Ladd approach converges sufficiently rapidly that it is expected to be applicable to systems with more complex interatomic potentials, or even within the framework of DFT-based MD simulations, provided large enough systems can be considered to account for the strain fields around the steps and adequate sampling of the phonon spectra. Once reference values have been computed by the Frenkel-Ladd approach, the thermodynamic-integration formalism developed in this work can be used to compute step free energies incorporating configurational and vibrational contributions on an equal footing. In general, such calculations require combinations of efficient interatomic potential models and/or advanced sampling methods to enable equilibration of kink and point-defect densities. Additionally, the contributions due to capillary fluctuations can give rise to large size effects (due to the long-wavelength modes) particularly near the roughening temperature (\eg, Refs.~\onlinecite{fisher} and \onlinecite{binder}), and to account for these effects calculations with different system sizes and/or analysis of the capillary wave spectra may be necessary. Nevertheless, provided these various considerations are taken into account, the formalism presented in this work provides a framework for computing benchmark results against which theories for vibrational (\eg, Refs.~\onlinecite{rahman} and \onlinecite{duruka}) and configurational (\eg, Refs.~\onlinecite{nelson} and \onlinecite{fisher}) contributions to the step free energies can be compared. We thus anticipate the approach to be useful for furthering understanding of the thermodynamic properties of steps on crystalline surfaces well beyond the application demonstrated in this paper.

  \begin{acknowledgments}
    The authors would like to thank Professor J.J. Hoyt for valuable discussions. The research of R.F. and M.A. at UC Berkeley was supported by the US National Science Foundation (Grants No. DMR-1105409 and No. DMR-1507033). R.F. acknowledges additional support from the Livermore Graduate Scholar Program. T.F. acknowledges partial support through a postdoctoral fellowship from the Miller Institute for Basic Research in Science at the University of California, Berkeley. Additional support for T.F. was provided under the auspices of the U.S. Department of Energy by Lawrence Livermore National Laboratory under Contract No. DE-AC52-07NA27344.
  \end{acknowledgments}

  \appendix*
  \section{\label{sec:appendix} Error calculation and numerical convergence analysis for the thermodynamic integration results}
    In this appendix we present analyses of the statistical sampling errors and numerical convergence for the thermodynamic integration results presented in Sec.~\ref{sec:results_TI}.
    
\begin{figure}
        \includegraphics[width=0.48\textwidth]{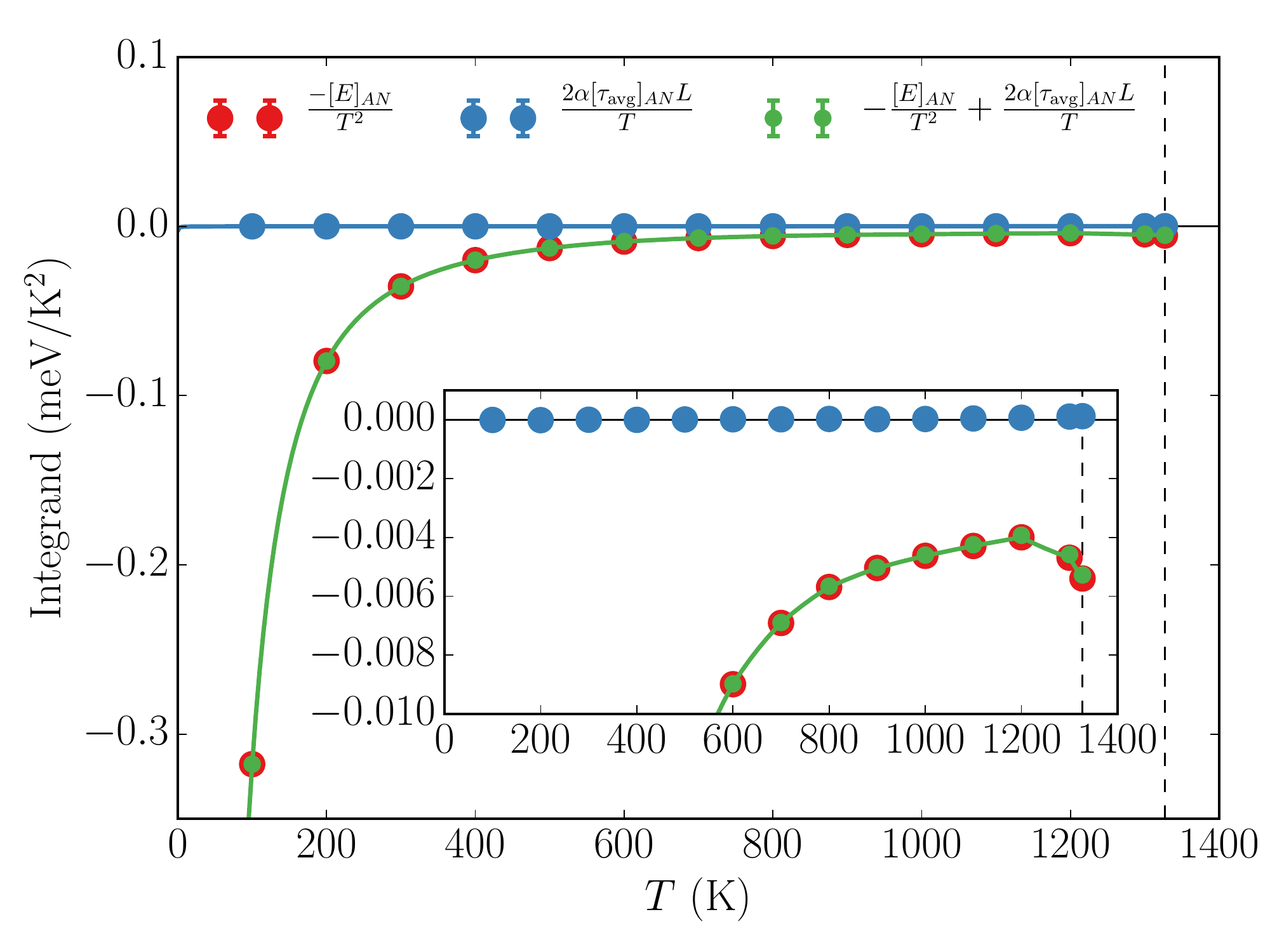}
        \caption{\label{fig:integrand} Temperature dependence of each term of the integrand of Eq.~\eqref{eq:TI_step}. Notice how the integrand term that involves the step excess stress $[\tau_\t{avg}]_{AN}$ is much smaller than the term involving the step excess energy $[E]_\t{AN}$. The solid lines are the result of the linear interpolation of the excess quantities multiplied by the factor of each term in the integrand.}
      \end{figure}

    The numerical integration in Eq.~\eqref{eq:TI_step} was performed considering a linear interpolation of the excess quantities shown in Fig.~\ref{fig:excess_vs_T}. We have tested interpolation schemes using polynomials of different orders and the difference compared to the linear interpolation was negligible. The reason is that the integrand of Eq.~\eqref{eq:TI_step} is already smooth for the linear interpolation due to the renormalization of the excess quantities by $T^2$ or $T$, as shown in Fig.~\ref{fig:integrand}. The second term in the integrand of Eq.~\eqref{eq:TI_step} (involving the excess stress) was found to be at least 50 times smaller than the first term (involving the excess energy) and therefore it is numerically negligible for the result of the integral.
    
\begin{figure}
        \includegraphics[width=0.48\textwidth]{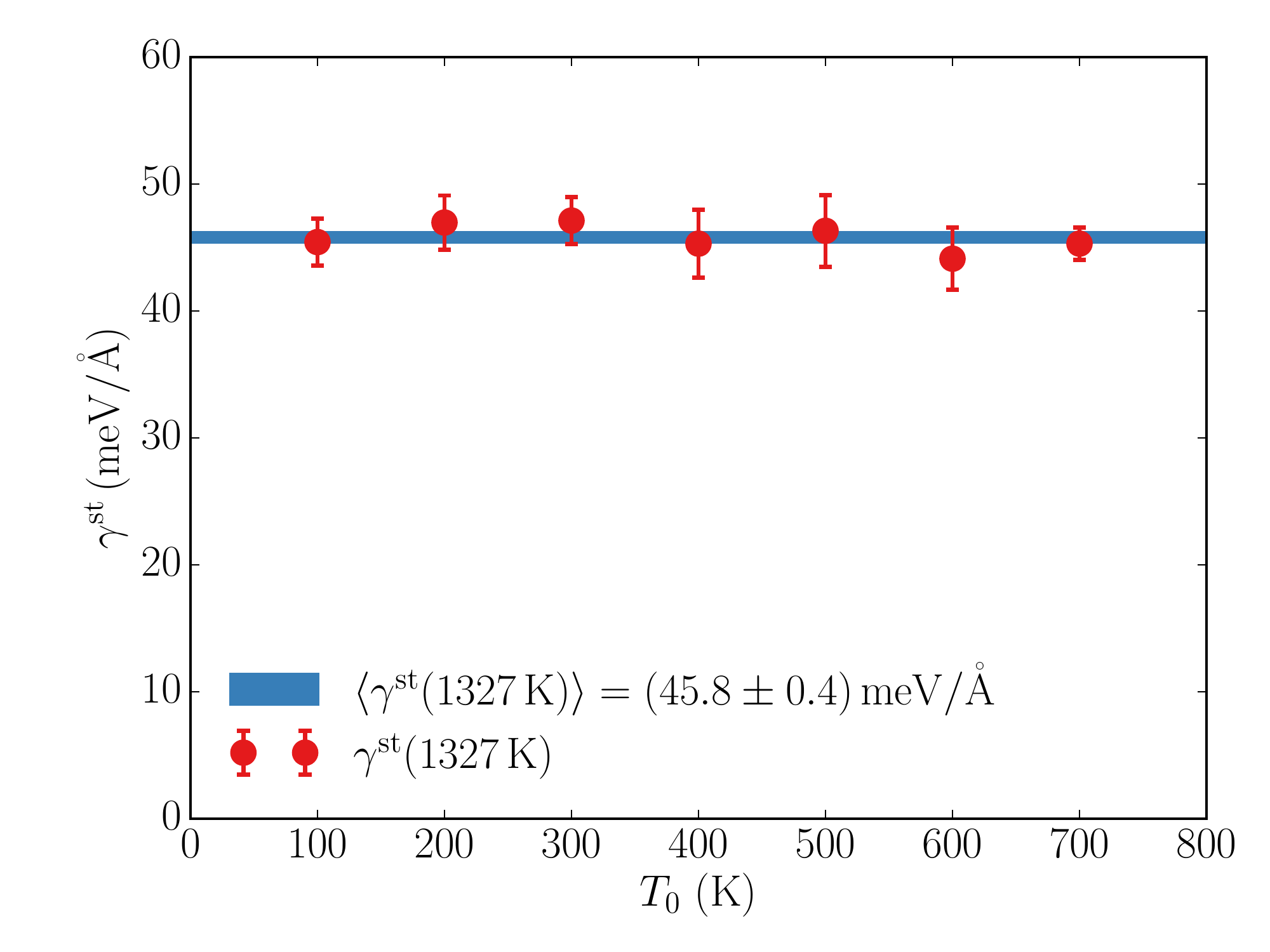}
        \caption{\label{fig:gamma_step_T0}Step free energy at $T_\t{m} = 1327\K$ calculated with the thermodynamic integration method using different initial points $\gamma^\t{st}(T_0)$ for the integration. The blue stripe is centered on the average taken considering all choices for $T_0$ from $100\K$ to $700\K$, and the width of this stripe corresponds to the standard deviation of these values for $\gamma^\t{st}(1327\K)$.}
      \end{figure}

\begin{figure}
        \includegraphics[width=0.48\textwidth]{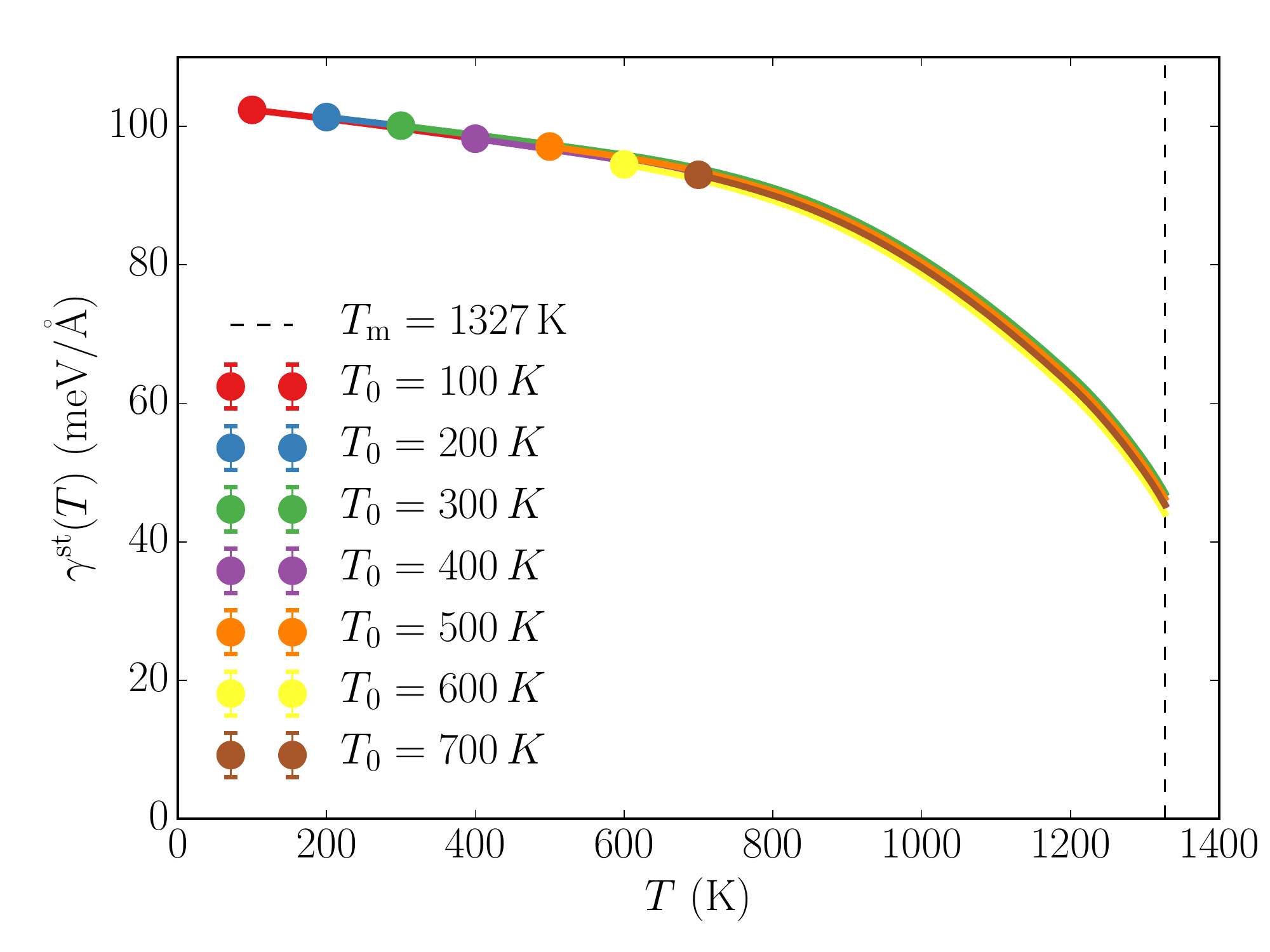}
        \caption{\label{fig:gamma_step_all_T0}Temperature dependence of the step free energy obtained using different reference temperatures $T_0$ for the thermodynamic integration [Eq.~\eqref{eq:TI_step}].}
      \end{figure}

      We have chosen $T_0 = 400\K$ as the initial point to perform the thermodynamic integrations to compute the temperature dependence of $\gamma^\t{st}$.  The integration was performed in both directions, from $T_0$ to $T_\t{m}$ and from $T_0$ to $\approx 0\K$. The choice of $T_0 = 400\K$ as the initial integration point in Fig.~\ref{fig:excess_vs_T} was arbitrary and, within the statistical accuracy of the calculations, it should not influence the final results for $\gamma^\t{st}$. The free energy calculated with the FL method at any of the other temperatures (red points in Fig.~\ref{fig:gamma_step}) should all be equally valid as an initial integration point. Thus, to verify the accuracy of the calculations, we have performed the integration in Eq.~\eqref{eq:TI_step} starting from all the different $T_0$ values for which we have available FL simulations. The result is shown in Fig.~\ref{fig:gamma_step_T0} where we plot the value of $\gamma^\t{st}$ at $T_\t{m} = 1327\K$ obtained from the integration of Eq.~\eqref{eq:TI_step} using different initial points. The error bar of each point corresponds to the error of the mean for the particular value of $T_0$.  The error in the mean was obtained by a resampling process of the excess quantities and the initial value of $\gamma^\t{st}$ used in the integration: each of the data points involved in the integration was picked randomly from a normal distribution with a mean value corresponding to the calculated average value of that quantity, and the standard deviation corresponding the calculated standard error of the mean value. The linear interpolation and numerical integration of the excess quantities for the given choice of $T_0$ was performed and the resulting step free energy at $T_\t{m}$ was averaged over $2000$ of these resampled data sets. For completeness we also show in Fig.~\ref{fig:gamma_step_all_T0} the $\gamma^\t{st}(T)$ curve obtained from the integration starting from the different $T_0$ values. From Figs.~\ref{fig:gamma_step_T0} and \ref{fig:gamma_step_all_T0} it is clear that the choice of the initial integration point $T_0$ does not influence the final result of the thermodynamic integration significantly.
    
  \newpage
  \bibliography{bibliography}
\end{document}